\documentclass{aa}
\usepackage[varg]{txfonts}
\usepackage{graphicx}
\usepackage[version=3]{mhchem} 
\usepackage{pifont} 
\usepackage{longtable}
\bibpunct{(}{)}{;}{a}{}{,} 
\usepackage{color}

\begin{document}

        \title{Modeling snowline locations in protostars: The impact of the structure of protostellar cloud cores}
        
        \author{N. M. Murillo\inst{1} \and T.-H. Hsieh\inst{2,3} \and C. Walsh\inst{4}}
        
        \institute{Star and Planet Formation Laboratory, RIKEN Cluster for Pioneering Research, Wako, Saitama 351-0198, Japan\\ \email{nadia.murillomejias@riken.jp}
                \and Max-Planck-Institut für extraterrestrische Physik, Giessenbachstrasse 1, 85748 Garching, Germany
                \and Academia Sinica Institute of Astronomy and Astrophysics, 11F of Astronomy-Mathmatics Building, AS/NTU, No.1, Section 4, Roosevelt Road, Taipei 10617, Taiwan
                \and School of Physics and Astronomy, University of Leeds, Leeds, LS2 9JT, UK}
        
        \abstract
        {Snowlines during star and disk formation are responsible for a range of effects during the evolution of protostars, such as setting the chemical composition of the envelope and disk. This in turn influences the formation of planets by changing the elemental compositions of solids and affecting the collisional properties and outcomes of dust grains. Snowlines can also reveal echoes of past accretion bursts, providing insight into the formation process of stars.}
        {The objective is to identify which parameters (e.g., luminosity, gas density, and presence of disk) dictate the location of snowlines during the early, deeply embedded phase and to quantify how each parameter changes the observed snowline location.}
        {A numerical chemical network coupled with a grid of cylindrical-symmetric physical models was used to identify what physical parameters alter the \ce{CO} and \ce{H2O} snowline locations. The investigated parameters are the initial molecular abundances, binding energies of \ce{CO} and \ce{H2O}, heating source, cloud core density, outflow cavity opening angle, and disk geometry. Simulated molecular line emission maps were used to quantify the change in the snowline location with each parameter.}
        {The snowline radius of molecules with low sublimation temperatures ($\lesssim$30 K), such as \ce{CO}, shift outward on the order of 10$^{3}$ AU with an order of magnitude increase in protostellar luminosity. An order of magnitude decrease in cloud core density also shifts the \ce{CO} snowline position outward by a few 10$^{3}$ AU. The presence of disk(-like) structures cause inward shifts by a factor of a few, and mainly along the disk mid-plane. For molecules that sublimate at higher temperatures, such as \ce{H2O}, increasing the protostellar luminosity or decreasing the cloud core density by an order of magnitude shifts the snowline position outward by a factor of a few. The presence of a disk concentrates molecules with high sublimation temperatures to compact regions (a few 10 AU) around the protostar by limiting the outward shift of snowline positions. Successful observational measurements of snowline locations are strongly dependent on spatial resolution, the presence or lack thereof of disk(-like) structures, and the inclination of the disk(-like) structure.}
        {The \ce{CO} and \ce{H2O} snowline locations do not occur at a single, well-defined temperature as is commonly assumed. Instead, the snowline position depends on luminosity, cloud core density, and whether a disk is present or not. Inclination and spatial resolution affect the observability and successful measurement of snowline locations. We note that \ce{N2H+} and \ce{HCO+} emission serve as good observational tracers of \ce{CO} and \ce{H2O} snowline locations. However, constraints on whether or not a disk is present, the observation of additional molecular tracers, and estimating envelope density will help in accurately determining the cause of the observed snowline position. Plots of the \ce{N2H+} and \ce{HCO+} peak emission radius versus luminosity are provided to compare the models with observations of deeply embedded protostars aiming to measure the \ce{CO} and \ce{H2O} snowline locations.} 
        
        
        \titlerunning{Modeling snowline locations in protostars}
        \authorrunning{Murillo, Hsieh \& Walsh}
        
        \maketitle
        
        \section{Introduction}
        \label{sec:intro}
        
        The physical structure of a cloud core determines the type of protostellar system that forms and how it evolves.
        The chemical structure within a cloud core, while an excellent tool to probe the physical structure, is also relevant in star and planet formation.
        The distribution and phase change of molecules as well as the resulting elemental composition of gas and ice decide the chemical complexity of the planets, comets, and meteorites that may form as part of the protostellar system.
        The region where a molecule undergoes a phase change to and from gas and ice is referred to as its snowline.
        Hence understanding how the physical structure determines the distribution of chemical content, including the snowline location, is relevant to understanding both the dynamical processes that occur during star formation as well as the chemical composition of planets, comets, and meteorites.
        
        Within a star-forming cloud core, the protostar is the main source of luminosity and heat due to the release of gravitational energy from contraction and material accretion.
        The amount of protostellar heating dictates the temperature structure of the cloud core.
        In an idealized spherical scenario, the temperature alone would dictate the snowline locations within the cloud core.
        However, star-forming cloud cores are not spherically symmetric.
        The outflow cavity, flattened structures around the protostar (e.g., pseudo-disks and rotationally supported disks; hereafter referred to as disks for simplicity), and variations within the envelope density can all impact how heat is distributed within the cloud core.
        Studies have shown that heating  mainly escapes through the outflow cavity in deeply embedded sources \citep{vankempen2009,yildiz2015,murillo2018b}.
        Thus the extent of chemical richness in the outflow cavity provides insight into the luminosity of the protostar and the physical conditions of the envelope (e.g., \citealt{drozdovskaya2015,murillo2018a,tychoniec2019,tychoniec2020}).
        Observations of embedded protostars (so-called Class 0 and I systems) have shown the presence of disks, both as flattened dust continuum structures (e.g., \citealt{jorgensen2009,enoch2011,persson2016,seguracox2018,tobin2020}) and rotationally supported disks traced in molecular gas (e.g., \citealt{murillo2013,harsono2014,yen2015,yen2017,maret2020}).
        These disks show a wide range of geometries and radii ranging from a few 10 AU up to $\sim$200 AU.
        Additional studies have shown that the presence of a disk can alter the temperature profile along the equator (disk mid-plane) of the cloud core (e.g., \citealt{murillo2015,murillo2018a,vanthoff2018b,hsieh2019a}).
        Multiplicity, that is two or more protostars within a single cloud core, can produce further asymmetries due to differences in luminosity from the multiple components and their locations with respect to each other (e.g., \citealt{chen2009,koumpia2016,murillo2016,murillo2018a}).
        At an early evolutionary stage, protostellar luminosity is dominated by accretion, that is accretion luminosity \citep{hartmann1996}.
        Variability in protostellar luminosity has been detected toward several targets (V1647 Ori: \citealt{abraham2004,andrews2004,acosta-pulido2007,fedele2007,aspin2009}; OO Serpentis: \citealt{kospal2007}; CTF93 216-2 \citealt{caratti-o-garatti2011}; VSX J205126.1: \citealt{covey2011,kospal2011}; HOPS383 \citealt{safron2015}; S255IR-SMA1 \citealt{caratti-o-garatti2016,liu2018}; EC53 \citealt{herczeg2017,yoo2017}).
        Such variability is considered to be a product of the nonuniform accretion of material with a variable amount and frequency onto the protostar, that is to say episodic accretion \citep{audard2014}.
        
        This variability can then change the chemical structure of the cloud core, leading to a dynamic chemical evolution \citep{taquet2016,molyarova2018}.
        Molecular snowlines provide a way to characterize one aspect of the chemistry within the cloud core by indicating where a particular molecule undergoes a phase change (from gas to ice and vice versa).
        Hence, the snowline of a molecular species is defined as the radius at which the species is half frozen onto the dust grains \citep{hayashi1981,vanthoff2017}. 
        Snowline locations are expected to shift back and forth during the star formation process due to the variable accretion luminosity of protostars (e.g., \citealt{hsieh2019b}).
        Thus, measuring snowline locations can provide insight into protostellar evolution during the early embedded phase.
        In addition, given the likelihood of planets forming during the early stages of star formation (e.g., \citealt{tychoniec2020}), these shifts likely also impact the formation and composition of planets.
        Whether planets form inside or outside of particular snowlines, such as the water snowline, influences the atmospheric C/O ratios and core compositions of planets (e.g., \citealt{oberg2011,walsh2015,eistrup2016,bosman2019}).
        
        Observational line emission images of protostars have contributions from different processes (infall, outflow, rotation, internal, and external heating) and structures (envelope, outflow cavity wall, outflow, jet, and disk), in addition to geometrical effects imposed by the viewing angle. 
        It is then difficult to determine what factors are producing the observed chemical structure. 
        Consequently, deducing the snowline location of different molecular species can be challenging.
        Hence, simple physico-chemical models that explore and compare the impact of individual parameters can provide insight as to what factors affect the observed chemical structure.
        In addition, because observed chemical signatures may reflect the thermal history of the protostellar system, the comparison of models and observations can reveal the occurrences of different processes. 
        For example, some chemical signatures are believed to trace a previous outburst of mass accretion \citep{lee2007,jorgensen2013,taquet2016,molyarova2018,wiebe2019}.
        
        Several models studying the location of snowlines have been previously published.
        Some of them focus on the snowline location of specific molecular species \citep{rab2017,vanthoff2017,frimann2017}, model a particular physical structure \citep{harsono2015,bjerkeli2016,rab2017}, have limited chemical modeling \citep{harsono2015,frimann2017,owen2020}, or consider more molecular species but use a spherically symmetric envelope without the addition of a disk or outflow cavity \citep{visser2015}.
        The effects of gas dispersal and dust evolution on the \ce{CO} snowline location has been studied in disks around Herbig stars \citep{panicmin2017}.
        The effect of dust grain sizes on molecular gas distributions and the \ce{H2O} snowline location has been modeled for a T Tauri disk \citep{gavino2021}.
        These models all provide important insight into the factors that dictate the chemical structure set by snowlines, but they are unable to more generally constrain the impact of outflow cavities and disk-like structures on the location of snowlines within the cloud core.

        This paper explores the physical conditions that affect the location of \ce{CO} and \ce{H2O} snowlines, as well as quantifies the resulting abundance distribution and emergent line emission from their respective tracers, \ce{N2H+} and \ce{HCO+}.
        Since \ce{CO} destroys \ce{N2H+} via an ion-molecule reaction (\cee{N2H+ + CO -> HCO+ + N2}), \ce{N2H+} increases in abundance when \ce{CO} freezes out onto the dust grains, thus tracing the \ce{CO} snowline location.
        A similar process occurs between \ce{H2O} and \ce{HCO+} (\cee{HCO+ + H2O -> H3O+ + CO}).
        To achieve this, a 2D (cylindrical symmetric) physical model was coupled with a reduced chemical network that includes key formation and destruction for each species. 
        With cylindrical symmetry, the disk, outflow cavity, and envelope structures were included.
        The conditions of the physical model were varied within a range of parameters in order to study the effects of different physical structures within the cloud core. 
        The range of parameters was chosen to reflect the variety of observed conditions in protostars.
        Section~\ref{sec:methods} describes the physical model, chemical network, and how simulated emission line maps were generated. 
        Section~\ref{sec:results} describes the results from the molecular distribution models and simulated emission maps, highlighting which parameters affect the snowline location and observability.
        The discussion and conclusions are presented in Sections~\ref{sec:discussion} and \ref{sec:conclusions}, respectively.
        
        \begin{table*}
                \centering
                \caption{Model parameters}
                \begin{tabular}{c c c c}
                        \hline \hline
                        Description & Parameter & Range & Units \\
                        \hline
                        \multicolumn{4}{c}{Physical}\\
                        \hline
                        Effective temperature & $T_{\rm eff}$ & 1500, 5000 & K \\
                        Central Stellar luminosity & $L_{\rm cen}$ & 0.01, 0.1, 1, 10, 20, 30, 40, 50, 75, 100, 200 & L$_{\odot}$ \\
                        Envelope density & $\rho_{\rm env, 0}$ & 10$^{5}$, 10$^{6}$, 10$^{7}$, 10$^{8}$ & cm$^{-3}$ \\
                        Outflow cavity opening angle & $\Theta_{\rm outflow}$ & 20, 50, 100 & Deg \\
                        Disk radius & $R_{\rm disk}$ & 0\tablefootmark{a}, 50, 150 & AU \\
                        Disk mass & $M_{\rm disk}$ & 0\tablefootmark{a}, 5$\times$10$^{-3}$, 5$\times$10$^{-2}$, 1.5$\times$10$^{-1}$ & M$_{\odot}$ \\
                        Disk scale height & $H_{\rm 0}$/$R_{\rm disk}$ & 0\tablefootmark{a}, 0.05, 0.1, 0.3 & Dimensionless ratio \\
                        \hline
                        \multicolumn{4}{c}{Chemical} \\
                        \hline
                        \ce{N2} binding energy & $E_{\rm b,\ce{N2}}$ & 955\tablefootmark{b} & K \\
                        \ce{CO} binding energy & $E_{\rm b,\ce{CO}}$ & 1150, 1307\tablefootmark{b} & K \\
                        \ce{H2O} binding energy & $E_{\rm b,\ce{H2O}}$ & 4820, 5700\tablefootmark{b} & K \\
                        \hline
                \end{tabular}
                \\
                \tablefoot{\tablefoottext{a}{For the models without a disk.}
                        \tablefoottext{b}{References: $E_{\rm b,\ce{N2}}$, \citealt{anderl2016}; $E_{\rm b,\ce{CO}}$, \citealt{collings2003}, \citealt{collings2004}, \citealt{bisschop2006}, and \citealt{noble2012}; and $E_{\rm b,\ce{H2O}}$, \citealt{sandford1993} and \citealt{fraser2001}.}}
                \label{tab:parameters}
        \end{table*}
        
        \begin{table}
                \centering
                \caption{Fiducial model parameters}
                \begin{tabular}{c c c}
                        \hline \hline
                        Parameter & Without disk & With disk \\
                        \hline
                        $T_{\rm eff}$ & \multicolumn{2}{c}{5000 K} \\
                        $L_{\rm cen}$ & \multicolumn{2}{c}{10 L$_{\odot}$} \\
                        $\rho_{\rm env}$ & \multicolumn{2}{c}{10$^{6}$} cm$^{-3}$ \\
                        $\Theta_{\rm outflow}$ & \multicolumn{2}{c}{50$^{\circ}$} \\
                        $R_{\rm disk}$ & -- & 150 AU \\
                        $M_{\rm disk}$ & -- & 0.05 M$_{\odot}$ \\
                        $H_{\rm 0}$/$R_{\rm disk}$ & -- & 0.05 \\
                        $E_{\rm b,\ce{N2}}$ & \multicolumn{2}{c}{955 K} \\
                        $E_{\rm b,\ce{CO}}$ & \multicolumn{2}{c}{1150 K} \\
                        $E_{\rm b,\ce{H2O}}$ & \multicolumn{2}{c}{4820 K} \\
                        \hline
                \end{tabular}
                \label{tab:fiducial}
        \end{table}

        \begin{table*}
                \centering
                \caption{Chemical network reactions and adopted rate coefficients}
                \begin{tabular}{c c c c c c}
                        \hline \hline
                        ID & Reaction & $\zeta$ & $\alpha$ & $\beta$ & $\gamma$ \\
                        &           & (s$^{-1}$) & (cm$^{3}$ s$^{-1}$) & & (K) \\
                        \hline
                        1       & H$_2^+$ + H$_2$ $\longrightarrow$ H$_3^+$ + H     & ...   & 2.08$\times10^{-9}$   & 0.00  & 0.0 \\
                        2       & H$_3^+$ + HD $\longrightarrow$ H$_2$D$^+$ + H$_2$ & ...   & 3.50$\times10^{-10}$  & 0.00  & 0.0 \\
                        3       & H$_2^+$ + e$^-$ $\longrightarrow$ H + H       & ...     & 1.60$\times10^{-9}$   & -0.43 & 0.0   \\
                        4       & H$_3^+$ + e$^-$ $\longrightarrow$ H$_2$ + H     & ...   & 6.70$\times10^{-8}$   & -0.52 & 0.0   \\
                        5       & H$_2$D$^+$ + e$^-$ $\longrightarrow$ H$_2$ + D     & ...   & 6.79$\times10^{-8}$   & -0.52 & 0.0 \\
                        6       & HCO$^+$ + e$^-$ $\longrightarrow$ CO + H       & ...   & 2.80$\times10^{-7}$   & -0.69 & 0.0 \\
                        7       & DCO$^+$ + e$^-$ $\longrightarrow$ CO + D       & ...   & 2.40$\times10^{-7}$   & -0.69 & 0.0 \\
                        8       & N$_2$H$^+$ + e$^-$ $\longrightarrow$ N$_2$ + H     & ...   & 2.60$\times10^{-7}$   & -0.84 & 0.0   \\
                        9       & N$_2$D$^+$ + e$^-$ $\longrightarrow$ N$_2$ + D     & ...   & 2.60$\times10^{-7}$   & -0.84 & 0.0   \\
                        10      & H$_3^+$ + CO $\longrightarrow$ HCO$^+$ + H$_2$ & ...   & 1.61$\times10^{-9}$   & 0.00  & 0.0 \\
                        11      & H$_2$D$^+$ + CO $\longrightarrow$ DCO$^+$ + H$_2$ & ...   & 5.37$\times10^{-10}$  & 0.00  & 0.0  \\
                        12      & H$_2$D$^+$ + CO $\longrightarrow$ HCO$^+$ + HD    & ...   & 1.07$\times10^{-9}$   & 0.00  & 0.0   \\
                        13      & N$_2$H$^+$ + CO $\longrightarrow$ HCO$^+$ + N$_2$ & ...   & 8.80$\times10^{-10}$  & 0.00  & 0.0   \\
                        14      & N$_2$D$^+$ + CO $\longrightarrow$ DCO$^+$ + N$_2$ & ...   & 8.80$\times10^{-10}$  & 0.00  & 0.0   \\
                        15      & H$_3^+$ + N$_2$ $\longrightarrow$ N$_2$H$^+$ + H$_2$ & ...   & 1.80$\times10^{-9}$   & 0.00  & 0.0   \\
                        16      & H$_2$D$^+$ + N$_2$ $\longrightarrow$ N$_2$D$^+$ + H$_2$ & ...   & 6.00$\times10^{-10}$  & 0.00  & 0.0   \\
                        17      & H$_2$D$^+$ + N$_2$ $\longrightarrow$ N$_2$H$^+$ + HD    & ...   & 1.20$\times10^{-9}$   & 0.00  & 0.0   \\
                        18      & HCO$^+$ + D $\longrightarrow$ DCO$^+$ + H       & ...   & 1.00$\times10^{-9}$   & 0.00  & 0.0   \\
                        19      & DCO$^+$ + H $\longrightarrow$ HCO$^+$ + D       & ...   & 2.20$\times10^{-9}$   & 0.00  & 796.0 \\
                        20      & H$_3^+$ + D $\longrightarrow$ H$_2$D$^+$ + H     & ...   & 1.00$\times10^{-9}$   & 0.00  & 0.0   \\
                        21      & H$_2$D$^+$ + H $\longrightarrow$ H$_3^+$ + D     & ...   & 2.00$\times10^{-9}$   & 0.00  & 632.0 \\
                        22      & N$_2$H$^+$ + D $\longrightarrow$ N$_2$D$^+$ + H     & ...   & 1.00$\times10^{-9}$   & 0.00  & 0.0   \\
                        23      & N$_2$D$^+$ + H $\longrightarrow$ N$_2$H$^+$ + D     & ...   & 2.20$\times10^{-9}$   & 0.00  & 550.0 \\
                        24      & H$_2$ + cr $\longrightarrow$ H$_2^+$ + e$^-$   & 1.21$\times10^{-17}$  & ...   & ...   & ...   \\
                        25      & H$_2$D$^+$ + p-H$_2$ $\longrightarrow$ H$_3^+$ + HD    & ...   & 1.40$\times10^{-10}$  & 0.00  & 232.0 \\
                        26      & H$_2$D$^+$ + o-H$_2$ $\longrightarrow$ H$_3^+$ + HD    & ...   & 7.00$\times10^{-11}$  & 0.00  & 61.5 \\
                        27  & \ce{H2O + HCO+ -> H3O+ + CO} & ... & 2.10$\times10^{-9}$ & -0.50 & 0.0 \\
                        28  & \ce{H3O+ + e- -> H2O + H} & ... & 1.10$\times10^{-7}$ & -0.50 & 0.0 \\
                        \hline
                \end{tabular}
                \label{tab:chemnetwork}
        \end{table*}
        
        \begin{figure}
                \centering
                \includegraphics[width=0.9\linewidth]{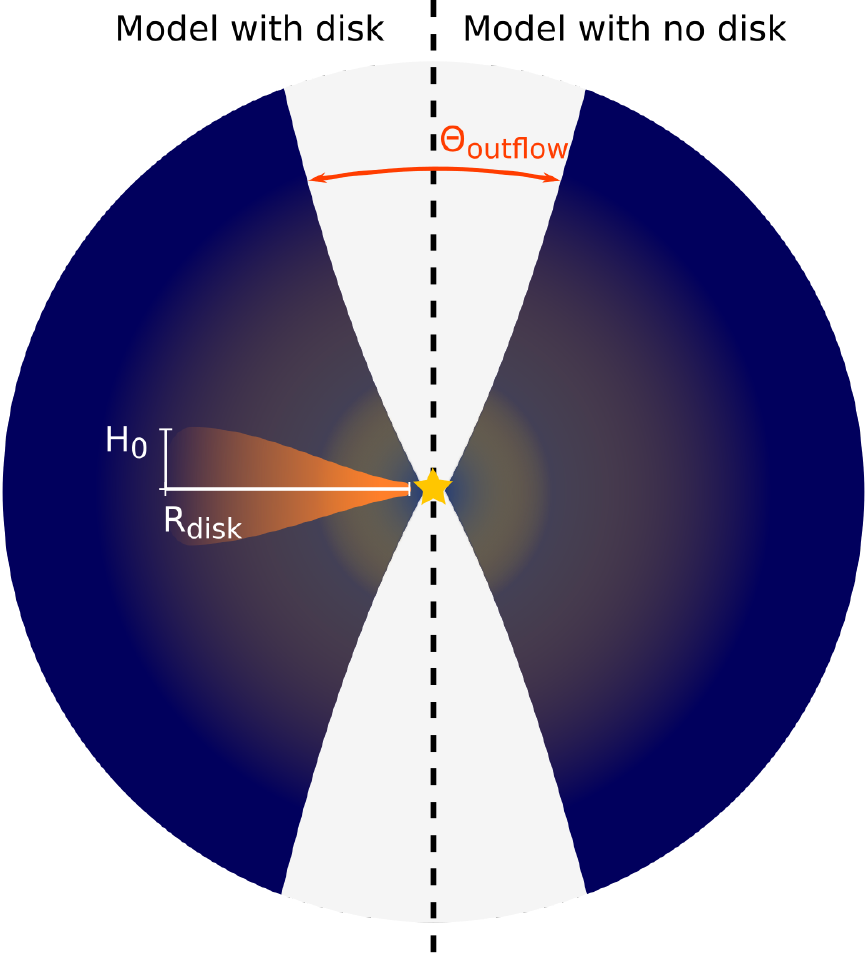}
                \caption{Cartoon showing the structure of the physical models, along with the illustrated definitions of the disk radius $R_{\rm disk}$, disk scale height $H_{0}$ at $R_{\rm disk}$, and outflow cavity opening angle $\Theta_{\rm outflow}$.} 
                \label{fig:cartoon}
        \end{figure}
        
        \begin{figure*}
                \centering
                \includegraphics[width=\textwidth]{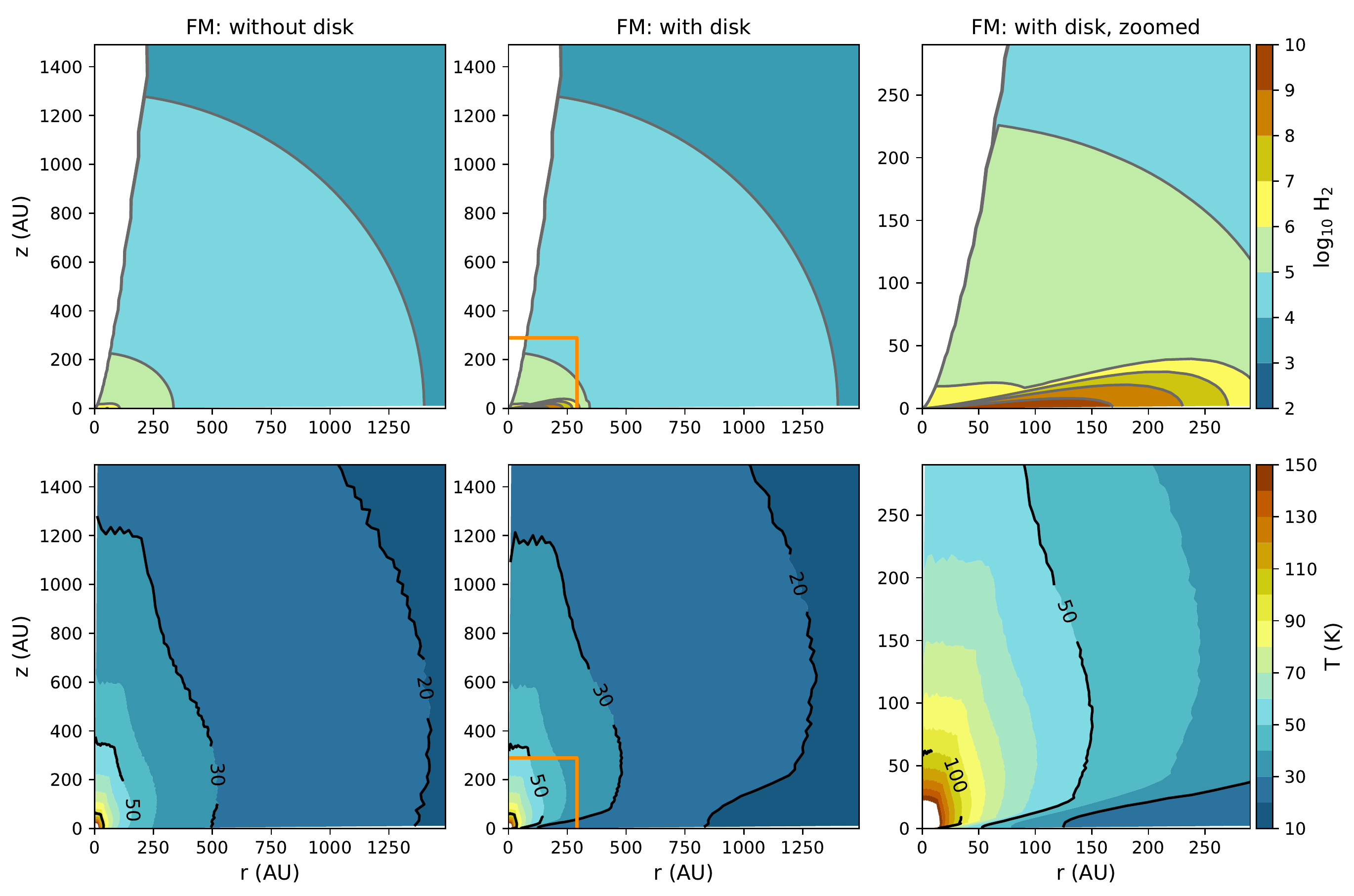}
                \caption{Density (\textit{top row}) and temperature (\textit{bottom row}) distribution of the two fiducial models: without a disk (\textit{left}) and with a disk (\textit{center and right}). The right column shows a zoom in of the fiducial model with the disk marked with an orange box in the center column.}
                \label{fig:fiducial}
        \end{figure*}
        
        \section{Methods}
        \label{sec:methods}
        
        \subsection{Physical models}
        \label{subsec:physsetup}
        Our physical models consist of a disk, outflow cavity, envelope, and a central heating source (also referred to as the protostar). 
        The structure is illustrated in Figure~\ref{fig:cartoon}.
        The density profile is built starting from a rotating flattened envelope introduced by \citet{ulrich1976}
        \begin{equation}
                \rho_{\rm env}(r,\mu)=\rho_{\rm env,0}\left(\frac{r}{r_{\rm cen}}\right)^{-1.5}~\left(1+\frac{\mu}{\mu_0}\right)^{-1/2}~\left(\frac{\mu}{2\mu_0}+\frac{r_{\rm cen}}{r}\mu_0^2\right)^{-1},
                \label{eq:envelope}
        \end{equation}
        where $r$ and $\theta$ are the spherical coordinates,
        $\rho_{\rm env, 0}$ is the density parameter used to scale the density profiles, $r_{\rm cen}$ is the centrifugal radius, and $\mu$ = $\cos(\theta)$. The parameter $\mu_0$ = $\cos(\theta_{0})$ is a solution of
        \begin{equation}
                \mu_0^3+\mu_0(r/r_{\rm cen}-1)-\mu(r/r_{\rm cen})=0
            \end{equation}
    and sets the location and direction of streamlines.
        The centrifugal radius $r_{\rm cen}$ is arbitrarily set to be 50 au in the models presented in this work, representative of embedded small disks.
        
        We adopted a density profile of a flared disk \citep{chiang1997,hartmann1998} as      
        \begin{equation}
                \rho_{\rm disk}(R,z)=\frac{\Sigma(R)}{\sqrt{2\pi}H(R)} \exp{\left[-\frac{1}{2}\left(\frac{z}{H(R)}\right)^2\right]}
                \label{eq:disk}
        \end{equation}
        with $\Sigma(R)$ defined as
        \begin{equation}
                \Sigma(R)=\Sigma_0\left(\frac{R}{R_{\rm disk}}\right)^{-1}\exp{\left[-\frac{R}{R_{\rm disk}}\right]}
        \end{equation}
        where $R$ and $z$ are the cylindrical coordinates, $R_{\rm disk}$ is the disk radius in au, and $\Sigma_0$ is the disk surface density at $R_{\rm disk}$.
        The disk scale height H(R) is defined as a function of radius R
        \begin{equation}
                H(R)=H_0\left(\frac{R}{R_{\rm disk}}\right)^{1.3},
                \label{eq:pressureheight}
        \end{equation}
        \citep{chiang1997}, for which $H_0$ is the pressure scale height at the radius $R_{\rm disk}$, and is tuned as a ratio normalized to $R_{\rm disk}$.
        
        The final gas density at a specific pixel is taken as the larger value between $\rho_{\rm env}$ and $\rho_{\rm disk}$.
        To convert the \ce{H2} density (cm$^{-3}$) to dust density (g~cm$^{-3}$), we assumed a gas-to-dust mass ratio of 100 and a mean molecular weight for the gas of 2.3.
        It is important to note that the dust evolution is not taken into account, and we assume that the dust and gas are well mixed.
        The dust opacity including absorption and scattering is obtained from the DIANA Opacity Tool\footnote{http://dianaproject.wp.st-andrews.ac.uk/data-results-downloads/fortran-package/} \citep{woitke2016}. This tool computes fast models given a grain size distribution.
        A power-law index of 3.5 ($dn/da \propto a^{-3.5}$, where $n$ is the number of grains with radius $a$) is used for both envelope and disk grain size distribution, with a maximum grain size of 1 $\mu$m, and 100 $\mu$m for the envelope, and disk, respectively.
        So whilst the full effects of dust evolution are not being taken into account, the assumption that dust grains in the disk are on average bigger than those in the envelope is adopted.
        
        An outflow cavity is carved out from the envelope defined by
        \begin{equation}
                z=z_0\times\left(\frac{R}{z_0\tan(\Theta_{\rm outflow}/2)}\right)^{1.5}
                \label{eq:outflow}
        \end{equation}
        where $\Theta_{\rm outflow}$ is the outflow cavity opening angle. We adopt ${\rm z_0\equiv50~au}$.
        
        Two-dimensional density and temperature profiles are generated with the above described structures.
        The dust temperature is calculated with RADMC3D\footnote{http://www.ita.uni-heidelberg.de/~dullemond/software/radmc-3d/} \citep{dullemond2012}, which uses the Monte Carlo method, for a given dust opacity and density with a central heating source. 
        An effective temperature $T_{\rm eff}$ and central stellar luminosity $L_{\rm cen}$ are set for the central heating source.
        In practice, the stellar radius is tuned to the luminosity using the Stefan-Boltzmann law with a given $T_{\rm eff}$.
        
        \subsubsection{Model grid setup}
        \label{subsec:grid}
        A grid of models is generated in order to study the effect of physical conditions on the locations of the \ce{CO} and \ce{H2O} snowlines within the cloud core of embedded protostellar sources.
        The parameters which characterize the protostar, outflow, disk, and envelope are treated as free parameters. 
        For each model one parameter is changed with the purpose to recognize the factors that influence the location of snowlines.
        Table~\ref{tab:parameters} lists the parameters along with the range used in the grid of models.
        
        The luminosity of the central protostar $L_{\rm cen}$ (L$_{\odot}$) is altered given the observed luminosity range of protostars and the variability produced by accretion processes. 
        In addition, the effective temperature of the star $T_{\rm eff}$ (K) is also kept as a free parameter, and can reflect the mass of the protostar, and its evolutionary stage.
        For example, $T_{\rm eff}$ = 1500 K is suitable for a proto-brown dwarf, whereas $T_{\rm eff}$ = 5000 K is more representative of a solar-type protostar.
        As the uv-visible light from the star is absorbed and reradiated back out by the circumstellar dust at a longer wavelength, a higher effective temperature for the same luminosity can heat up the disk-envelope more efficiently.
        
        Photons from the protostar escape mainly through the outflow cavity, heating the gas and kickstarting chemical processes within the cavity walls (e.g., \citealt{drozdovskaya2015,yildiz2015,vankempen2009,murillo2018a,murillo2018b}). 
        It is of interest to test whether the outflow cavity opening angle has an effect on the location of snowlines within the cloud core.
        Observations suggest that the outflow cavity opening angle $\Theta_{\rm outflow}$ (degrees) increases as the protostar evolves \citep{arce2006,velusamy2014,hsieh2017,heimsoth2021}, while a recent study finds no evidence of outflow cavity opening angle growth during the Class I phase \citep{habel2021}. 
        However, these studies suffer from a lack of good evolutionary indicators \citep{whitney2003,offner2009}, and the outflow opening angle can be different depending on the tracers used to make the measurement.
        The outflow opening angle is defined in Equation~\ref{eq:outflow}.
        It should be noted, however, that the models presented here only include simple molecular species. 
        Modeling of complex organic molecules along the outflow cavity walls has been previously studied using more comprehensive gas-grain chemistry in, for example, \citet{drozdovskaya2015}.
        
        As material is accreted from the cloud core onto the protostar (and the disk), the envelope density $\rho_{\rm env}$ decreases with time, thus changing this parameter emulates the evolution of the cloud core.
        The envelope density $\rho_{\rm env}$ is defined in Eq.~\ref{eq:envelope}. 
        
        Studies in the last decade have revealed both the presence of and lack of disks at all stages of protostellar evolution, with a wide range of physical characteristics (Reviews: \citealt{williams2011,belloche2013,li2014}; Surveys: \citealt{yen2015,testi2016,yen2017,maury2019,tobin2020}; Individual sources: \citealt{murillo2013,aso2017,lee2017,hsieh2019a}).
        Hence, in our physical model, the outer disk radii $R_{\rm disk}$ (AU), total disk mass $M_{\rm disk}$ (M$_{\odot}$), and disk scale height $H_{0}$ are treated as free parameters.
        The disk density is constrained by these three parameters. 
        Thus, for two disks with the same mass, one with a large radius and scale height will be less dense than one with a small radius and scale height. 
        Models without disks are also generated in order to further explore the effect of these disk structures on the location of snowlines.
        In the case of no disk, the density structure is simply described by the flattened envelope profile (see Eq.~\ref{eq:envelope}).

        \subsubsection{Fiducial models}
        As a point of reference, two fiducial models are generated: one model without a disk, and one with a disk.
        The density and temperature distributions of the fiducial models are shown in Fig.~\ref{fig:fiducial}, and all parameters for the fiducial models are listed in Table~\ref{tab:fiducial}.

        Both fiducial models have an effective stellar temperature of $T_{\rm eff}$ = 5000 K, and envelope density of $\rho_{\rm env}$ = 10$^{6}$ cm$^{-3}$, representative of solar type stars.
        A relatively high luminosity of $L_{\rm star}$ = 10 L$_{\odot}$ is chosen to better highlight features for comparison. 
        Although not all protostellar sources exhibit such high luminosities, some embedded sources have been observed to have bolometric luminosities on the order of 10 L$_{\odot}$ (e.g., \citealt{murillo2016}).
        The outflow cavity opening angle is set to $\Theta_{\rm outflow}$ = 50$^{\circ}$ based on observations of embedded protostars \citep{arce2006}.
        
        Observations have shown the presence of large ($\geq$100 AU) disk(-like) structures around protostars (e.g., \citealt{murillo2013,harsono2014,persson2016,tobin2018}), while further surveys show that smaller disk(-like) structures are common (e.g., \citealt{harsono2014,yen2015,yen2017,maury2019,maret2020}).
        The fiducial model with a disk is then chosen to have a disk with $R_{\rm disk}$ = 150 AU, $M_{\rm disk}$ = 0.05 M$_{\odot}$, and $H_{\rm 0}$ =  0.05. 
        The disk mass is chosen based on an assumed central protostar mass of 0.5 M$_{\odot}$.
        This is reasonable assuming the central protostar is a solar type star ($T_{\rm eff}$ = 5000 K) that has accreted at least half its mass in the embedded phase.
        
        \subsection{Chemical network}
        \label{subsec:chem}
        
        Given the grid size of each physical model, and the total number of physical models, for speed and simplicity, we use a reduced chemical network based on that compiled by the UMIST Database for Astrochemistry \citep{woodall2007,mcelroy2013}. 
        Our reduced network captures the main gas-phase reactions important for the formation of \ce{HCO+} and \ce{N2H+}, and associated deuterated ions such as \ce{H2D+}, \ce{DCO+}, and \ce{N2D+} (see Table~\ref{tab:chemnetwork}). 
        As we are interested in \ce{HCO+} as an accessible tracer of the \ce{H2O} snowline location, and \ce{N2H+} and \ce{DCO+} as tracers of the \ce{CO} snowline position, we only consider singly deuterated forms, and neglect spin-state chemistry for simplicity. 
        We allow the freeze-out of molecules onto dust grain surfaces as well as thermal desorption, and photodesorption (induced by both cosmic rays and stellar photons). 
        We assume that \ce{H2O}, \ce{CO} and \ce{N2} have already formed. 
        It is assumed that water is already present in the cloud core, in either ice or gas forms (\citealt{vanDishoeck2013,vanDishoeck2017}, and references therein).
        Additional reactions include the destruction of water by reaction with \ce{HCO+}, and the formation of water through the dissociative recombination of \ce{H3O+}.
        These two reactions are used to model the water abundance in a simple way, with the key reaction destroying gas-phase water being \ce{H2O + HCO+}. 
        The dissociative recombination of \ce{H3O+} is included to reform the gas-phase water and is necessary to close the network.
        We do not include grain-surface chemistry, but we do allow the recombination of gas-phase cations with negatively charged grains. 
        We note that this approach will not accurately model the chemistry occurring in the outflow cavity walls; however, we are interested in the locations of snowlines, which are present only in well-shielded gas.
        The reactions included in the network, along with the parameters to calculate the rate coefficients of each reaction are listed in Table~\ref{tab:chemnetwork}.
        Our treatment of the chemistry represents an intermediate choice between a parametric model (e.g., \citealt{yildiz2010}) and a full chemical model (e.g., \citealt{drozdovskaya2015,notsu2021}). Future work will explore the impact of a more complex chemical network on the abundances and distribution of key snowline tracers.
        
        Several parameters of the chemical network can be changed in order to determine if a resulting outcome is due to a chemical or physical effect.
        These parameters include the initial abundances of the species in the network, and the binding energies $E_{\rm b}$ of \ce{CO}, \ce{N2}, and \ce{H2O} (Table~\ref{tab:parameters}).
        Initial abundances are set for \ce{CO}, \ce{N2}, \ce{H2O}, \ce{HD}, and neutral grains, and naturally, for \ce{H2} and \ce{H}.
        The total initial abundances of \ce{CO}, \ce{N2}, \ce{H2O}, \ce{HD} and neutral grains relative to \ce{H2} are set to 2$\times$10$^{-4}$, 1$\times$10$^{-4}$, 2$\times$10$^{-4}$, 1.6$\times$10$^{-5}$, and 1.3$\times$10$^{-12}$, respectively.
        The initial abundances for \ce{CO}, \ce{N2} and \ce{H2O} are set to all ice.
        The binding energy value reflects the conditions under which a molecule is bound to grain surfaces.
        The binding energy of \ce{N2} is kept constant for all models, and is set at $E_{\rm b,\ce{N2}}$ = 955 K \citep{collings2004,garrod2006}.
        To probe molecular binding under different ice environments and determine how binding energy affects the \ce{CO} and water snowline locations, two binding energies are used for each molecule: $E_{\rm b,\ce{CO}}$ = 1150 K and 1307 K for \ce{CO} \citep{collings2003,collings2004,bisschop2006,noble2012}; $E_{\rm b,\ce{H2O}}$ = 4820 K and 5700 K for water \citep{sandford1993,fraser2001}. 
        To simulate the absorption and desorption processes, we use an average dust grain size of $0.1\mu$m. 
        Thus, absorption and desorption rates will be determined by the temperature and gas density (with a fixed gas-to-dust ratio) at a given binding energy (e.g., \citealt{anderl2016}).
        
        The chemical network is evolved up to 10$^{8}$ years to ensure that steady state has been reached to produce molecular distributions for each set of conditions.
        Given that timescales for protostellar evolution and disk formation are shorter than this, we checked the validity of this assumption by comparing the results at 10$^{8}$ years with those at earlier times. We can confirm that steady state in our reduced network is already reached by 1 Myr timescales.
        Because of this, there is no difference if the initial abundances are set to all ice, all gas, or a mix.
        While the network has the capability of time-dependent chemical models, in this paper the steady state output is used since the aim is to study the effect of cloud core physical structures on the locations of the \ce{CO} and water snowlines.
        Additional time-dependent models to study the effect of episodic accretion bursts will be considered in a subsequent paper.
        
        \begin{table*}
                \centering
                \caption{Transitions used for the simulated line emission maps}
                \begin{tabular}{c c c c c c c}
                        \hline \hline
                        Molecule & Transition & Frequency (GHz) & $E_{\rm up}$ (K) & Telescope & Convolving Beam & Reference\tablefootmark{a} \\
                        \hline
                        \ce{^{12}CO} & 2--1 & 230.5380 & 16.6 & ALMA Band 6 & 2$\arcsec$ & \citealt{hsieh2019b} \\
                        \ce{^{13}CO} & 2--1 & 220.3987 & 15.9 & ALMA Band 6 & 2$\arcsec$ & \citealt{hsieh2018} \\
                        \ce{C^{18}O} & 2--1 & 219.5604 & 15.8 & ALMA Band 6 & 2$\arcsec$ & \citealt{hsieh2018} \\
                        \ce{DCO+} & 3--2 & 216.1126 & 20.7 & ALMA Band 6 & 2$\arcsec$ & \citealt{murillo2018a} \\
                        \ce{N2H+} & 1--0 & 93.1734 & 4.5 & ALMA Band 3 & 2$\arcsec$ & \citealt{hsieh2019b} \\
                        \ce{p-H2O} & 3$_{\rm 1,3}$ -- 2$_{\rm 0,2}$ & 183.3101 & 204.7 & ALMA Band 5 & 0.25$\arcsec$ & \\
                        \ce{o-H2O} & 1$_{\rm 1,0}$ -- 1$_{\rm 0,1}$ & 556.9360 & 61.0 & Herschel HIFI & 0.25$\arcsec$\tablefootmark{b} & \citealt{kristensen2012} \\
                        \ce{HCO+} & 3--2 & 267.5576 & 25.7 & ALMA Band 6 & 0.25$\arcsec$ & \citealt{hsieh2019b} \\
                        \hline
                \end{tabular}
                \\
                \tablefoot{
                        \tablefoottext{a}{Previous work that reports observations of the transitions used for the line emission radiative transfer models.}
                        \tablefoottext{b}{The spatial resolution in \citet{kristensen2012} for \ce{o-H2O} is 39$\arcsec$. A much smaller beam of 0.25$\arcsec$ is used in order to compare with \ce{HCO+}.}}
                \label{tab:RTtrans}
        \end{table*}
        
        \subsection{Simulated line emission maps}
        \label{subsec:RT}
        
        In order to examine whether the effects of physical and chemical structure on the \ce{CO} and \ce{H2O} snowline locations are observable, simulated emission maps are generated from the molecular distributions. 
        The molecules \ce{N2H+} and \ce{HCO+} are typically used in observations to determine the \ce{CO} and \ce{H2O} snowline positions, respectively \citep{vanthoff2017,frimann2017,hsieh2018,hsieh2019b}.
        Simulated emission maps of \ce{CO} (three isotopologues: \ce{^{12}CO}, \ce{^{13}CO}, and \ce{C^{18}O}), \ce{N2H+}, \ce{DCO+}, \ce{H2O}, and \ce{HCO+} for the two fiducial models are produced in order to examine the robustness of the emission from these species in tracing snowline locations.
        Simulated emission maps of \ce{N2H+} and \ce{HCO+} for all the models discussed in Section~\ref{sec:results} are produced. These are used to determine which effects have an observable, and distinguishable, impact on the observation of snowline locations.
        
        The simulated line emission channel maps are calculated using RADMC3D under the assumption of local thermal equilibrium (LTE). 
        Seven different inclination angles are used: 0$^\circ$ (face-on), 15$^\circ$, 25$^\circ$, 45$^\circ$, 65$^\circ$, 75$^\circ$, and 90$^\circ$ (edge-on).
        The distance to Perseus of 293 pc \citep{ortiz2018,zucker2018} is adopted for ray-tracing and a map size of 60$\arcsec~\times$60$\arcsec$ (17580 $\times$ 17580 AU) is used for \ce{N2H+}, while a smaller map size of 6$\arcsec~\times$6$\arcsec$ (1758 $\times$ 1758 AU) is used for \ce{HCO+}. 
        Table~\ref{tab:RTtrans} lists the transitions modeled.
        These transitions are selected because they are commonly observed, and observations of these transitions toward protostars have been reported in literature.
        
        Observational parameters such as bolometric luminosity change with the inclination of the cloud core relative to the line of sight (e.g., \citealt{whitney2003,crapsi2008}).
        Thus, the spectral energy distribution (SED) of each model is also generated with RADMC3D for each inclination angle. 
        The modeled SEDs are used to calculate the bolometric luminosity after radiative transfer.
        
        To calculate the emergent line emission maps, a velocity field is required. 
        The velocity structures from infalling-rotating cores \citep{ulrich1976,terebey1984} are used, and defined as
        \begin{equation}
                V_r=-\left(\frac{GM}{r}\right)^{1/2}~\left(1+\frac{\cos\theta}{\cos{\theta_0}}\right)^{1/2},
        \end{equation}
        \begin{equation}
                V_\theta=\left(\frac{GM}{r}\right)^{1/2}~(\cos\theta_0-\cos\theta)~ \left(\frac{\cos\theta_0+\cos\theta}{\cos\theta_0\sin^2\theta}\right)^{1/2},
        \end{equation}
        \begin{equation}
                V_\phi=\left(\frac{GM}{r}\right)^{1/2}~\left(1-\frac{\cos\theta}{\cos\theta_0}\right)^{1/2}~\frac{\sin\theta_0}{\sin\theta}
        \end{equation}
        where $G$ is the gravitational constant, $M$ is the central source mass, $r$ is the radius, $\theta$ is the angle in polar coordinates, and $\theta_0$ is the angle of the initial velocity in polar coordinates.
        
        Keplerian rotation is adopted for the inner disk region with $R<r_{\rm cen}$. The gravitational force decreases as the orthogonal distance from the disk mid-plane increases such that \citep{pinte2018}
        \begin{equation}
                V_\phi=\left(\frac{GM}{R}\right)^{1/2} \sin(\theta),~r<r_{\rm cen}.
        \end{equation}
        For the radial and meridional velocities, a dimensionless factor is applied to the above equations as a function of $\theta$ and $r$, $(\frac{r}{r_{\rm cen}})^{2\theta/\pi}$. This factor gradually decreases the rotation velocity along the mid-plane, and, with a fixed $r_{\rm cen}$, the velocity field is scaled by the squared root of the central mass $M$ that is set to be $0.5~M_\odot$. 
        
        The modeled transitions and convolving beams from the Atacama Large Millimeter/submillimeter Array (ALMA) observations of \citet{hsieh2019b} are adopted. 
        For \ce{N2H+} 1-0 (ALMA Band 3), the convolving beam is 2$\arcsec$, while for \ce{HCO+} 3-2 (ALMA Band 6) the convolving beam is 0.25$\arcsec$.
        For water, two transitions with low upper energy $E_{\rm up}$ were used, one at 183 GHz with $E_{\rm up}$ = 205 K (ALMA Band 5), and a second at 557 GHz with $E_{\rm up}$ = 61 K (Herschel Space Observatory).
        The latter transition has been observed and reported in \citet{kristensen2012} for embedded Class 0 and I low-mass protostars.
        Continuum subtraction is not performed on the simulated line emission channel maps.
        
        \begin{table*}
                \centering
                \caption{Effect of each parameter on \ce{CO} and \ce{H2O} snowline locations}
                \begin{tabular}{c c c c c}
                        \hline \hline
                        Parameter & \multicolumn{2}{c}{\ce{CO} snowline} & \multicolumn{2}{c}{\ce{H2O} snowline} \\
                         & Molecular distribution & Simulated emission & Molecular distribution & Simulated emission \\
                        \hline
                        \multicolumn{5}{c}{Chemical Effects} \\
                        \hline
                        Binding Energy & \ding{51} & \ding{51}\tablefootmark{a} & \ding{51} & \ding{51}\tablefootmark{a} \\
                        \hline
                        \multicolumn{5}{c}{Physical Effects} \\
                        \hline
                        Effective Temperature $T_{\rm eff}$ & \ding{55} & \ding{55} & \ding{55} & \ding{55} \\
                        Luminosity $L_{\rm star}$ & \ding{51} & \ding{51} & \ding{51} & \ding{51} \\
                        Envelope Density $\rho_{env}$ & \ding{51} & \ding{51} & \ding{51} & \ding{51} \\
                        Outflow cavity opening angle $\Theta_{\rm outflow}$ & \ding{55} & \ding{55} & \ding{55} & \ding{55} \\
                        \hline
                        \multicolumn{5}{c}{Disk geometry} \\
                        \hline
                        Radius $R_{\rm disk}$ & \ding{51} & \ding{51} & \ding{55}\tablefootmark{b} & \ding{51}\tablefootmark{b} \\
                        Mass $M_{\rm disk}$ & \ding{51} & \ding{51} & \ding{55}\tablefootmark{b} & \ding{51}\tablefootmark{b} \\
                        Scale height $H_0$ & \ding{51} & \ding{51} & \ding{55}\tablefootmark{b} & \ding{55}\tablefootmark{b} \\
                        Density $\rho_{\rm disk}$ & \ding{51} & \ding{51} & \ding{55}\tablefootmark{b} & \ding{51}\tablefootmark{b} \\
                        \hline
                        \multicolumn{5}{c}{Observational Effects} \\
                        \hline
                        Spatial resolution & & \ding{55} & & \ding{51} \\
                        Inclination $i$ & & \ding{55} & & \ding{51} \\
                        \hline
                \end{tabular}
                \\
                \tablefoot{
                        \tablefoottext{a}{High spatial resolution is needed to detect the shift, and to differentiate it from other effects.}
                        \tablefoottext{b}{The presence of a disk in the cloud core, and its geometry, constrain the extent of the region where water is in the gas phase, particularly in the radial direction. In the molecular distribution models the presence of the disk does not significantly change the water snowline location. However, the geometry of the disk affects the observability and measurability of the \ce{H2O} snowline location.}}
                \label{tab:snowparam}
        \end{table*}

        \begin{figure*}
                \centering
                \includegraphics[width=\linewidth]{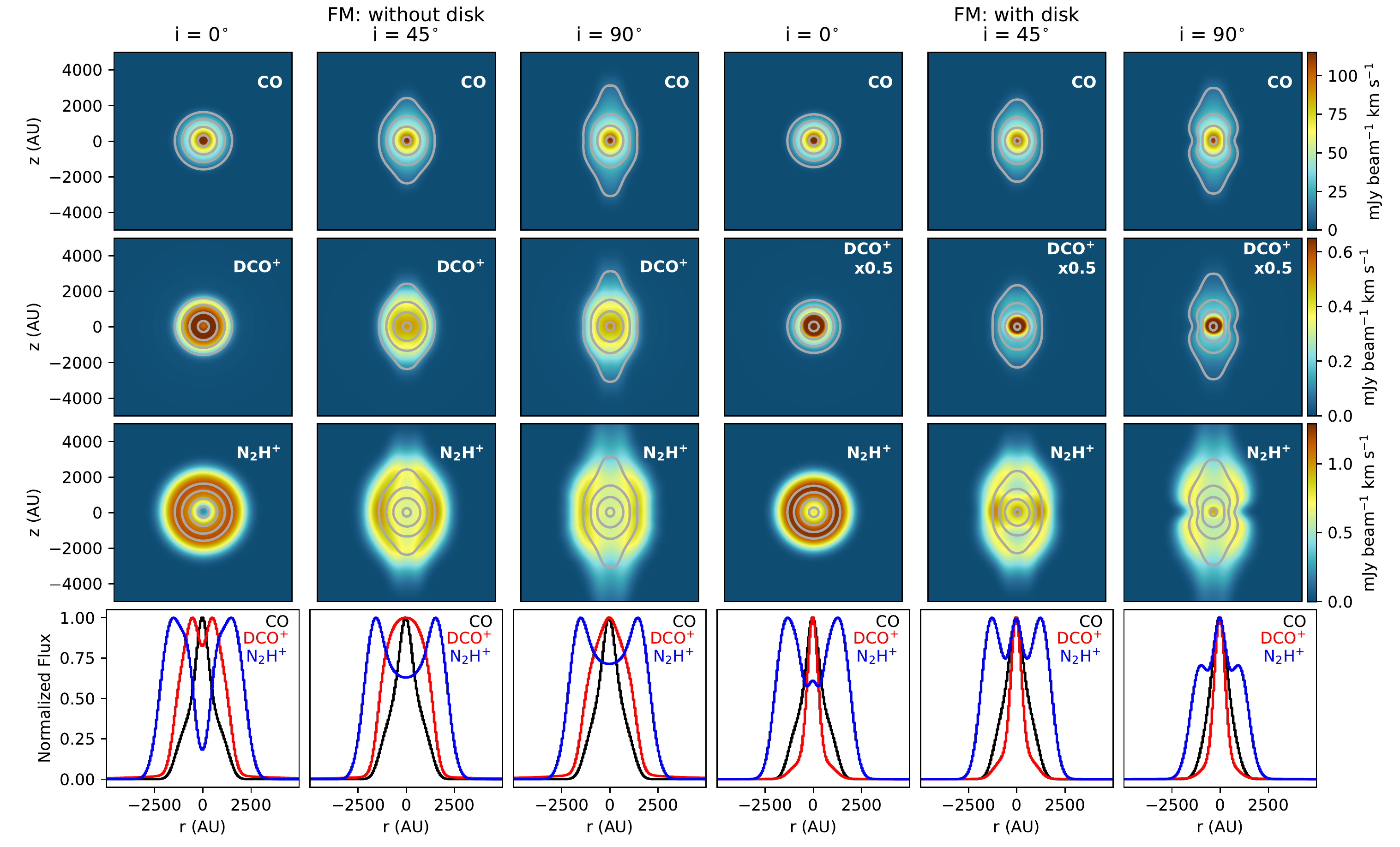}
                \caption{Robustness of \ce{N2H+} and \ce{DCO+} as \ce{CO} snowline location tracers. Integrated intensity maps of the simulated line emission show \ce{CO} (\textit{top row}), \ce{DCO+} (\textit{second row}) and \ce{N2H+} (\textit{third row}) emission. Contours in all three rows are for \ce{CO} at 10, 30, 50, and 100 mJy~beam$^{-1}$ km~s$^{-1}$. Images that have been scaled for better comparison have the scaling factor noted on the top right corner. Different inclinations are shown in order to determine if inclination affects the robustness of \ce{N2H+} and \ce{DCO+} as snowline position tracers. Inclinations shown are from face-on ($i = 0^{\circ}$) to edge-on ($i = 90^{\circ}$). The first three columns show the fiducial model without disk, while the other three columns show the fiducial model with disk. Slices extracted along $z = 0$ from the simulated line emission maps are shown in the bottom row normalized to the peak of each profile.}
                \label{fig:COsnowlineRT}
        \end{figure*}
        
        \begin{figure*}
                \centering
                \includegraphics[width=\linewidth]{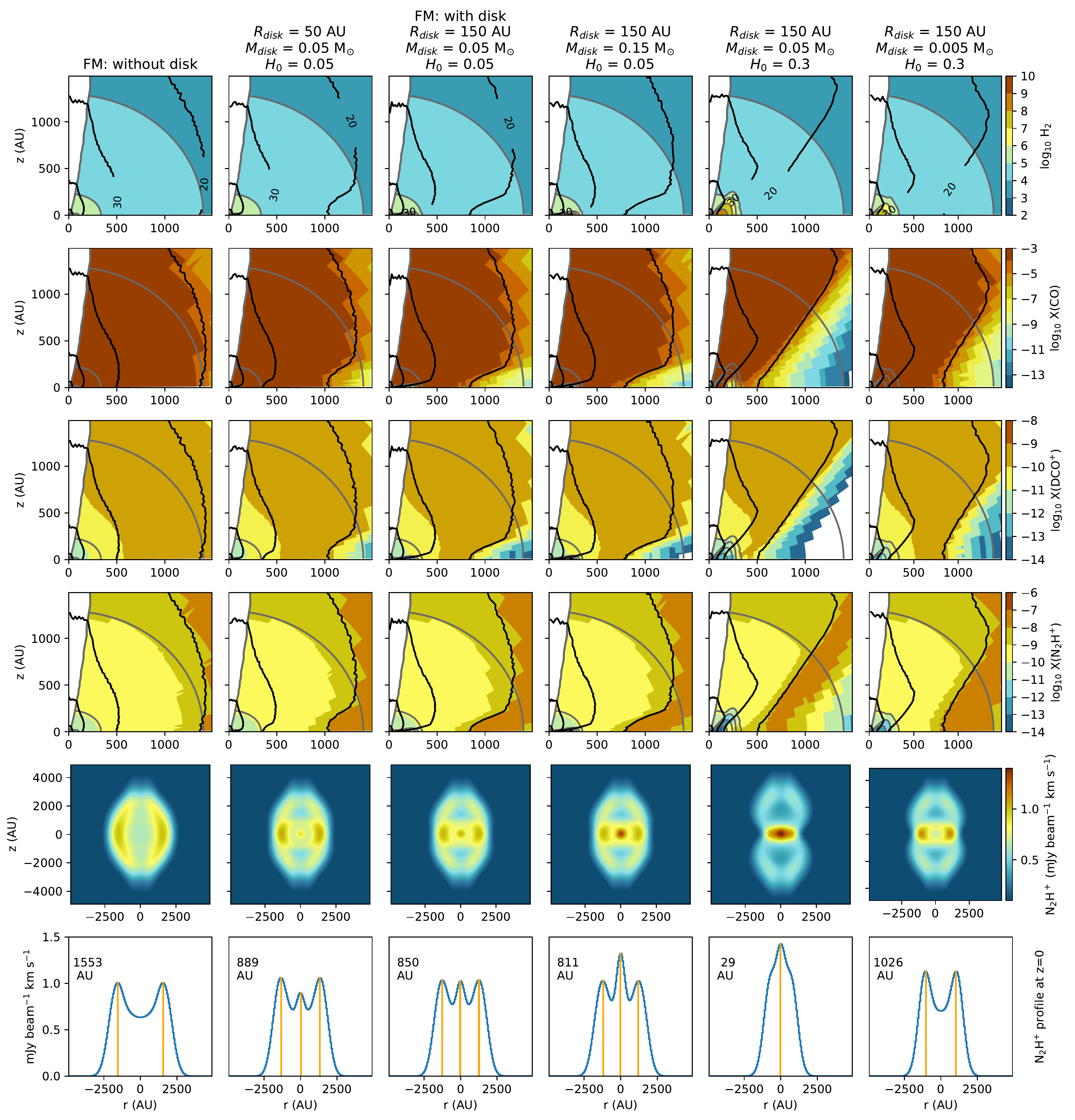}
                \caption{Effect of disk geometry parameters on the \ce{CO} snowline location. Molecular distributions are shown in the top four rows. The fifth row shows the maps of the integrated intensity of the simulated line emission for \ce{N2H+} 1--0. The sixth row shows the corresponding slice extracted along $z = 0$ from the \ce{N2H+} simulated emission maps. The average distance of the peaks from the center (i.e., peak radius) is indicated in AU in the top left corner. These positions are shown with orange vertical lines. The first and third columns show the fiducial models without and with disk, respectively. The additional columns show the effect of changing disk radius $R_{\rm disk}$ (\textit{second column}), disk mass $M_{\rm disk}$ (\textit{fourth column}), scale height $H_{0}$ (\textit{fifth column}), and disk density (\textit{sixth column}). Fractional abundances of all molecular species are relative to total number density of \ce{H2} (\textit{top row}). The black and gray contours show gas temperature and density, respectively. The emission maps are shown at $i$ = 45$^{\circ}$ and convolved to a beam of 2$\arcsec$.} 
                \label{fig:COdiskmodel}
        \end{figure*}
        
        \section{Results}
        \label{sec:results}
        The outcome of the models is described in this section.
        First, the robustness of each snowline location tracer is discussed based on the two fiducial models.
        Then each parameter explored in the models and their effect on each snowline position will be described. 
        In this way which parameters produce an observable effect can be examined.
        Table~\ref{tab:snowparam} lists the parameters studied and which snowline location is affected by a change in a parameter.
        Details on how the simulated line emission maps are generated are presented in Section~\ref{subsec:RT}.
        The effect of inclination and spatial resolution on the observed \ce{CO} and \ce{H2O} snowline locations will then be considered.
        Finally, plots of peak radius of \ce{N2H+} and \ce{HCO+} versus luminosity for the explored parameter space are provided and described.

        A sample of the models from the full grid is shown in Figures~\ref{fig:COdiskmodel} and \ref{fig:H2Odiskmodel}, as well as in Appendix A.
        The full grid of molecular distributions and simulated line emission maps are available online\footnote{The full model grid can be found at https://starformation.space/models/snowline-models/ and are also archived at CDS.}.

        \subsection{\ce{CO} snowline location}
        \label{subsec:COline}
        The two fiducial models are used to explore the robustness of \ce{N2H+} and \ce{DCO+} as tracers of the \ce{CO} snowline location.
        Figure~\ref{fig:COsnowlineRT} shows the integrated intensity (moment 0) maps of the simulated line emission, and slices extracted along $z = 0$.
        The slices are normalized to their respective peaks for comparison purposes, since \ce{CO} is several orders of magnitude brighter than \ce{N2H+} and \ce{DCO+}.
        From the $z = 0$ slices, \ce{N2H+} begins to increase at a radius where \ce{CO} is between 60\% and 70\% of its maximum brightness. The \ce{N2H+} emission then peaks where \ce{CO} is at $\sim$1\% of its maximum brightness.
        At low envelope densities (e.g., 10$^{5}$ cm$^{-3}$), \ce{N2H+} is no longer a robust tracer of the \ce{CO} snowline location (see Sect.~\ref{subsec:radvsparam}).
        In contrast, the \ce{DCO+} emission is located between the \ce{CO} and \ce{N2H+} peaks.
        From the $z = 0$ slice the \ce{DCO+} emission peaks before \ce{CO} has decreased to half its maximum brightness. 
        This is clearly visible in the case without a disk. 
        In the case with a disk, the \ce{DCO+} emission is less bright, but the trend is the same as evidenced by the \ce{DCO+} "wings" at $\sim$20\% brightness (Fig.~\ref{fig:COsnowlineRT}). 
        It should be noted that the central \ce{DCO+} peak in the case with a disk is a contribution from the continuum.
        Both behaviors, that of \ce{N2H+} and \ce{DCO+} are expected from the formation and destruction pathways of each molecule (see Table~\ref{tab:chemnetwork}, reactions 11 and 13).
        These results suggest that \ce{N2H+} and \ce{DCO+} can provide observational constraints on the \ce{CO} snowline location. 
        It should be noted, however, that the limited chemical network used in this work does not consider additional direct or indirect reactions leading to the formation or destruction of \ce{N2H+} and \ce{DCO+}. For example, the warm formation pathway for \ce{DCO+} (e.g., \citealt{favre2015,murillo2018a}) is not included in our current network.
        This is further discussed in Section~\ref{subsec:diss_robust}.
        Figure~\ref{fig:COiso_snowlineRT} shows how the emission from the \ce{CO} isotopologs, \ce{^{13}CO} and \ce{C^{18}O}, does not coincide well with the snowline location traced by the \ce{N2H+} emission.
        
        A radial shift of the \ce{CO} snowline location occurs throughout the envelope with changes in the binding energy $E_{\rm b,\ce{CO}}$, luminosity, and envelope density.
        These parameters change the \ce{CO} snowline position whether a disk is present or not.
        The location of the \ce{CO} snowline moves to smaller radii with an increase in the binding energy ($E_{\rm b,\ce{CO}}$ = 1150 $\rightarrow$ 1307 K, Fig.~\ref{fig:BE}).
        These binding energies result in sublimation temperatures of $\sim$21 K and $\sim$25 K, respectively.
        The simulated emission line maps show that the snowline position shifts by at least a few 100 AU when changing the binding energy, whether there is a disk present or not.
        Hence, the effect of $E_{\rm b,CO}$ on the \ce{CO} snowline location is potentially observable and measurable, but needs to be differentiated from other effects.
        With increasing luminosity $L_{\rm star}$, the 20 K gas temperature contour moves to larger radii, and consequently, so does the \ce{CO} snowline position.
        This is expected, and is widely used as a test for occurrence of episodic accretion in embedded protostellar systems.
        The gas density of the cloud core plays a major role in the distribution of different molecules.
        Thus, at low densities ($\rho_{\rm env}$ < 10$^{6}$ cm$^{-3}$) the \ce{CO} snowline location moves outward to larger radii, whereas at higher densities ($\rho_{\rm env}$ > 10$^{6}$ cm$^{-3}$) the snowline location moves inward to smaller radii (Fig.~\ref{fig:density}). 
        In either case, the snowline position shift does not coincide with the change in temperature structure. 
        This is due to the balance between the thermal desorption and freeze-out rates.
        The thermal desorption rate varies with density as n$^{b}$ where b is a value between 0 and 1, whereas the rate of freeze-out goes as $n^{2}$ \citep{cuppen2017}. 
        Thus, at low densities, the freezeout of molecules is slow causing snowline locations to occur at lower temperatures, and vice versa.
        Consequently, the snowline position does not occur at a single, well defined temperature as commonly assumed, but is also dependent upon density.
        In turn, density also affects the robustness of snowline tracers.
        The simulated line emission maps confirm that the change of cloud core density produces an observable and measurable effect on the snowline location.
        
        The presence of a disk structure produces an inward shift in the \ce{CO} snowline position along the disk mid-plane.
        From the molecular distributions (Fig.~\ref{fig:COdiskmodel}), it can be noted that ``rings'' of \ce{N2H+} and \ce{DCO+} are generated around the disk edge, in contrast to the more spherical distributions of both molecules in the case without a disk.
        For \ce{N2H+}, the effect is also noticeable in the simulated emission maps. For \ce{DCO+} the effect is seen in the simulated emission maps at $i$ = 0$^{\circ}$ (face-on), but less noticeable at $i$ $>$ 0$^{\circ}$ due to the contribution from the continuum component in the simulated emission maps. However, \ce{DCO+} rings have been observed toward embedded protostars (e.g., \citealt{murillo2018a}) and disks (e.g.,\citealt{mathews2013,salinas2018}). 
        The effect on the \ce{CO} snowline location from each parameter that defines the disk geometry is explored here.
        This is done by comparing the fiducial models with models where only one parameter is changed, either disk radius $R_{\rm disk}$, disk mass $M_{\rm disk}$, or scale height $H_{0}$.
        Figure~\ref{fig:COdiskmodel} shows the two fiducial models in addition to one model for each disk parameter that is changed. 
        The molecular distributions show that the 20 K gas temperature contour for the fiducial model without a disk is located at $\sim$1400 AU (Fig.~\ref{fig:COdiskmodel} first column).
        While a disk with $R_{\rm disk}$ = 50 and 150 AU cause the 20 K contour to move inward to 1000 and 800 AU along the disk mid-plane, respectively (Fig.~\ref{fig:COdiskmodel} second and third columns).
        Changing $M_{\rm disk}$ by a factor of a few or even an order of magnitude only shifts the \ce{CO} snowline location by $\sim$100 AU or less (Fig.~\ref{fig:COdiskmodel} third, fourth and sixth columns).
        In contrast to the molecular distributions, the simulated line emission maps do not show a particularly strong effect on the \ce{CO} snowline position as traced by \ce{N2H+} due to disk mass $M_{disk}$ and radius $R_{\rm disk}$ (Fig.~\ref{fig:COdiskmodel} bottom rows).
        Interestingly, changing $H_{0}$ not only alters the \ce{CO} snowline position along the disk mid-plane, but also vertically (Fig.~\ref{fig:COdiskmodel} third and fifth columns).
        This effect is evident in the simulated line emission images, where a flared disk (e.g., $H_{0}$ = 0.3) efficiently shadows the envelope both along the disk midplane and vertically.
        Thus, a flared disk causes a larger radial and vertical shift inward of the \ce{CO} snowline position (between 100 to 1000 AU) than that caused by a flatter disk.
        The effect of disk density on the \ce{CO} snowline location is implicit in the models (Fig.~\ref{fig:COdiskmodel} fifth and sixth columns).
        In the simulated line emission maps, disk density does not produce a significant snowline position shift, similar to disk mass and radius. 
        
        Two parameters in our models are found to have no effect on the \ce{CO} snowline location: effective temperature of the protostar, and outflow cavity opening angle.
        The effective temperature of the central protostar generates a more noticeable effect along the outflow cavity than within the cloud core.
        Thus, only a slight increase of the presence of \ce{CO} in the gas phase along the outflow cavity at z $>$ 1500 AU is apparent in the models.
        The outflow cavity opening angle $\Theta_{\rm out}$ only alters the \ce{CO} gas distribution at $r,z \gtrsim$ 1500 AU.

        \begin{figure*}
                \centering
                \includegraphics[width=\linewidth]{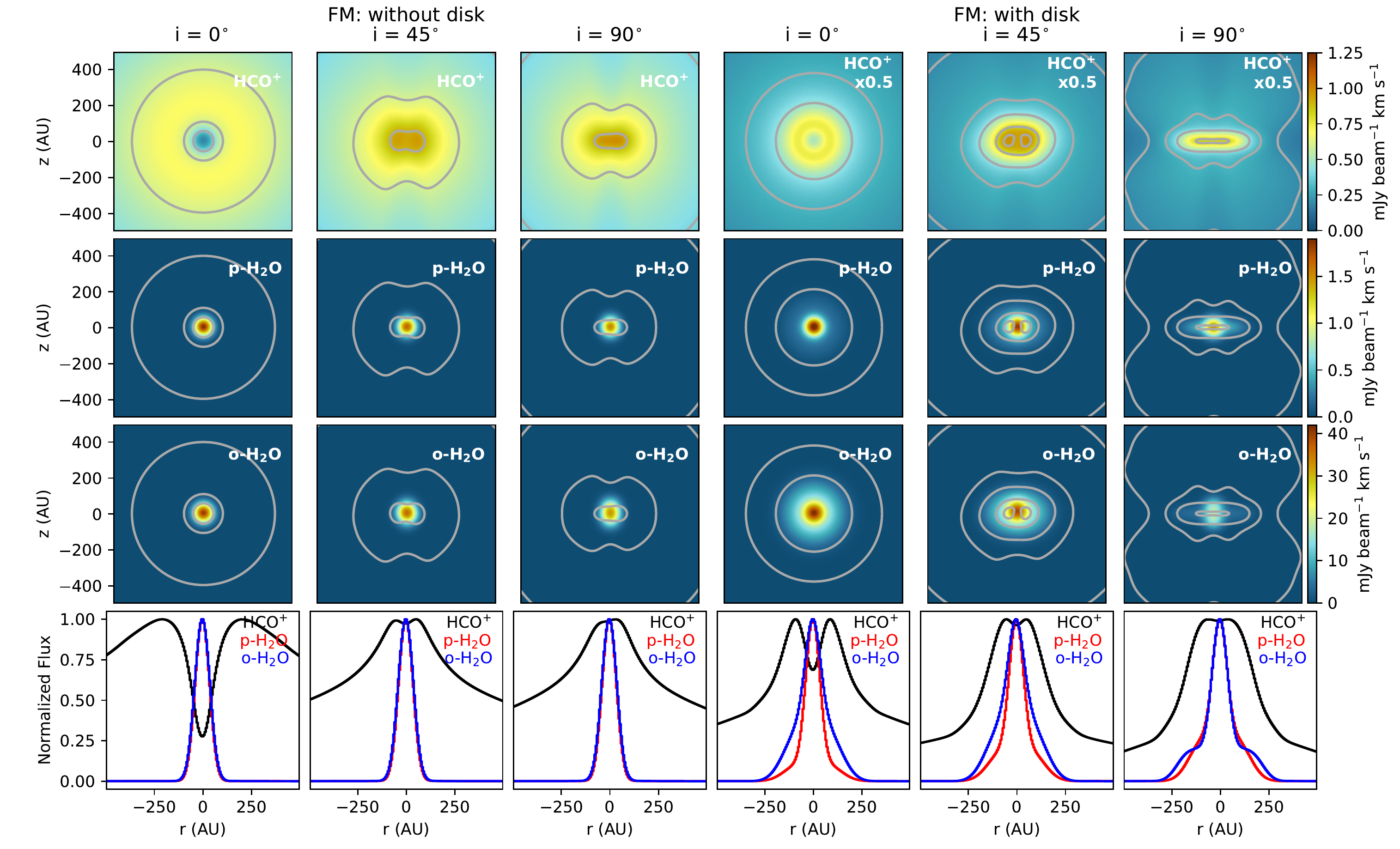}
                \caption{Robustness of \ce{HCO+} as a \ce{H2O} snowline location tracer. Simulated emission maps show \ce{HCO+} (\textit{top row}), \ce{p-H2O} (\textit{second row}) and \ce{o-H2O} (\textit{third row}) emission. Contours in all three rows are \ce{H2O} at 0.4, 0.6, 0.9, 1.6, 1.9, and 2.1 mJy~beam$^{-1}$ km~s$^{-1}$. Images that have been scaled for better comparison have the scaling factor on the top right corner. Different inclination are shown in order to determine if it affects the robustness of \ce{HCO+} as a snowline location tracer. Inclinations shown are from face-on ($i = 0^{\circ}$) to edge-on ($i = 90^{\circ}$). The first three columns show the fiducial model without disk, while the other three columns show the fiducial model with disk. Slices extracted along $z = 0$ from the simulated line emission maps are shown in the bottom row normalized to the peak of each profile.}
                \label{fig:H2OsnowlineRT}
        \end{figure*}
        
        \begin{figure*}
                \centering
                \includegraphics[width=\linewidth]{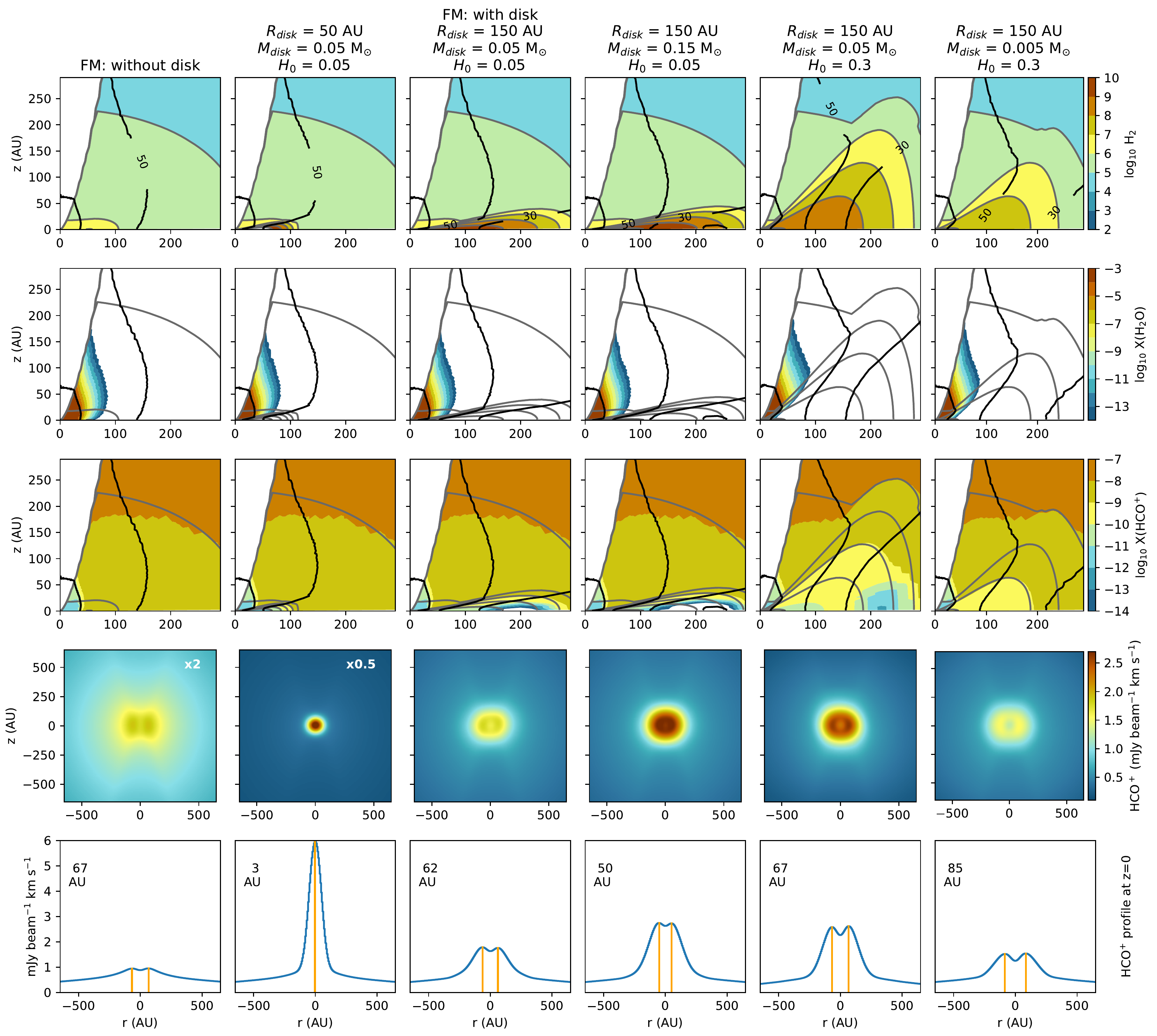}
                \caption{Effect of disk geometry parameters on the \ce{H2O} snowline location. Molecular distributions are shown in the top three rows. The fourth row shows the simulated intensity integrated line emission maps for \ce{HCO+} 3--2. Images that have been scaled for better comparison have the scaling factor on the top right corner. The fifth row shows the corresponding slice extracted along $z = 0$ from the \ce{HCO+} simulated emission maps. The average distance of the peaks from the center (i.e., peak radius) is indicated in AU in the top left corner. These positions are shown with orange vertical lines. No scaling has been applied to the \ce{HCO+} profiles. The first and third columns show the fiducial models without and with disk, respectively. The additional columns show the effect of changing disk radius $R_{\rm disk}$ (\textit{second column}), disk mass $M_{\rm disk}$ (\textit{fourth column}), scale height $H_{0}$ (\textit{fifth column}), and disk density (\textit{sixth column}). Fractional abundances of all molecular species are relative to total number density of \ce{H2} (\textit{top row}). The black and gray contours show gas temperature and density, respectively. The simulated line emission maps are shown at $i$ = 45$^{\circ}$ and convolved to a beam of 0.25$\arcsec$.}
                \label{fig:H2Odiskmodel}
        \end{figure*}
        
        \subsection{\ce{H2O} snowline}
        \label{subsec:waterline}
        
        The robustness of \ce{HCO+} emission as a water snowline location tracer is examined by using the two fiducial models (Fig.~\ref{fig:H2OsnowlineRT}).
        Models of both \ce{o-H2O} and \ce{p-H2O} emission are examined.
        \ce{HCO+} emission begins to increase at the \ce{H2O} half maximum, and peaks 20 to 40 AU further out.
        Similar to \ce{N2H+} for the \ce{CO} snowline, the \ce{HCO+} emission peak is not at the exact location of the water snowline, but provides an indication of where water begins to freeze-out onto the dust grains.
        As noted in Section~\ref{subsec:COline}, the limited chemical network in this work does not account for all the destruction and formation pathways for \ce{HCO+} and \ce{H2O}, in addition to the simple treatment of water in the network.
        
        The \ce{H2O} snowline location is altered by the binding energy $E_{\rm b,\ce{H2O}}$, luminosity, and envelope density.
        Increasing the binding energy of water ($E_{\rm b,\ce{H2O}}$ = 4820 $\rightarrow$ 5700 K) causes the \ce{H2O} snowline position to shift inward between 20 to 50 AU (Fig.~\ref{fig:BEH2O}). 
        The simulated emission line maps show that the effect of binding energy on the \ce{H2O} snowline location is observable.
        With sufficient spatial resolution and additional constraints, the effect can potentially be measured.
        Increasing the protostellar luminosity moves the 100 K gas temperature contour outward to larger radii, in turn shifting the location of the \ce{H2O} snowline as well.
        This is expected, given how much more luminous protostellar sources have relatively larger warm ($\gtrsim$100 K) regions than less luminous sources.
        Changing the envelope density has the same effect on the \ce{H2O} snowline location as it does on the \ce{CO} snowline position (Fig.~\ref{fig:densityH2O}).
        In the simulated emission line maps without a disk, the \ce{H2O} snowline location shifts proportional to luminosity, and inversely proportional to envelope density $\rho_{\rm env}$.
        Overall, the \ce{H2O} snowline location shifts on the order of $\sim$100 AU or less.
        
        Based on the molecular distributions, changing the disk radius $R_{\rm disk}$, disk mass $M_{\rm disk}$, and scale height $H_{0}$ has no apparent effect on the actual location of the \ce{H2O} snowline (Fig.~\ref{fig:H2Odiskmodel}).
        In a similar manner, the disk density does not seem to have a significant effect on the \ce{H2O} snowline location in the molecular distributions.
        However, the disk geometry does help to concentrate the \ce{H2O} gas extent between the outflow cavity wall and disk surface (Fig.~\ref{fig:H2Odiskmodel} fifth and sixth columns).
        Hence, a very flared disk ($H_{0}$ = 0.3) causes the gas-phase water to be located in a narrow region between the outflow cavity wall and the disk surface.
        Similarly, changing the outflow cavity opening angle will affect the vertical distribution of water gas but not the snowline location.
        This effect becomes relevant in the face-on inclination for the case without a disk (See Sect.~\ref{subsec:postproc}).
        The simulated line emission maps provide further insight into the \ce{H2O} snowline location when a disk is present (Fig.~\ref{fig:H2Odiskmodel} bottom rows).
        The presence of a disk concentrates the gas-phase water by limiting snowline location from shifting outward.
        This is somewhat dependent on luminosity and density (Sect.~\ref{subsec:radvsparam}), but occurs regardless of disk radius $R_{\rm disk}$, mass $M_{\rm disk}$, scale height $H_{0}$.
        
        The effective temperature of the central source only has a slight effect on the presence of gas-phase water in the z direction along the outflow cavity walls.
        The simulated line emission images further support this result.
        
        \subsection{Inclination and spatial resolution}
        \label{subsec:postproc}
        
        To properly compare the results presented in this work with observations, the inclination of the protostellar cloud core and the spatial resolution of the observations must be considered.
        The inclination $i$ of a cloud core, the orientation with respect to the line of sight, can change the observed source SED, derived bolometric luminosity, and determine which features of the circumstellar material at scales $<$100 AU can be observed (e.g., \citealt{crapsi2008}).
        The spatial resolution of observations is relevant for characterizing snowline locations, in particular those located within a few hundred AU.
        
        Figures~\ref{fig:COsnowlineRT} and \ref{fig:H2OsnowlineRT} show how inclination affects the peak location of \ce{N2H+} and \ce{HCO+} line emission, respectively, for the two fiducial models.
        In this work, $i$ = 0$^{\circ}$ is defined as face-on, whereas $i$ = 90$^{\circ}$ is edge-on.
        The peak position of simulated \ce{N2H+} emission decreases with increasing inclination (Fig.~\ref{fig:inclination}). This trend is more evident when a disk is present (Figs.~\ref{fig:LcenLbol_vspeak} to \ref{fig:disk_vspeak}).
        For $i$ = 0$^{\circ}$ to $i$ = 90$^{\circ}$, the peaks of \ce{N2H+} emission shift less than 500 AU in the case without a disk.
        In the case with a disk, the \ce{N2H+} emission peaks move inward from $i$ = 0$^{\circ}$ to $i$ = 90$^{\circ}$ by 25\%. 
        Hence, the inclination of the cloud core along the line of sight does not affect the robustness of \ce{N2H+} and \ce{DCO+} as \ce{CO} snowline location tracers.
        
        The simulated \ce{HCO+} emission peak radius decreases with increasing inclination (Fig.~\ref{fig:inclination}).
        In the case without a disk, the outflow cavity opening angle $\Theta_{\rm out}$ causes the \ce{HCO+} peak position to be overestimated for low inclinations and in particular $i$ = 0$^{\circ}$ (Fig.~\ref{fig:outflow_vspeak}). This is because $\Theta_{\rm out}$ alters the vertical distribution of the \ce{HCO+} gas, and looking down along the outflow cavity does not allow the radial and vertical distributions to be distinguished.
        When a disk is present, the disk itself is the main constraining factor and the \ce{HCO+} emission peak position is not overestimated at low inclinations due to $\Theta_{\rm out}$ or other factors.
        For the fiducial models, the peak moves from 210 ($i$ = 0$^{\circ}$) to 38 AU ($i$ = 90$^{\circ}$) for the case without a disk, and from 90 ($i$ = 0$^{\circ}$) to 9 AU ($i$ = 75$^{\circ}$) in the case with a disk (Fig.~\ref{fig:inclination}).
        At 0$^{\circ}$ $<~i~<$ 45$^{\circ}$, the \ce{HCO+} gap can be more clearly seen than at higher inclinations.
        At $i$ = 45$^{\circ}$, the \ce{HCO+} gap is only marginally seen, and may be difficult to observe depending on the disk geometry when a disk is present.
        This is most likely caused by the optical depth, that is, column density, of the \ce{HCO+} emission (see also \citealt{hsieh2019b,vanthoff2021}). 
        Thus, the robustness of \ce{HCO+} as a snowline position tracer of \ce{H2O} is affected by inclination, more so when a disk is present.
        Consequently, the \ce{H2O} snowline location can be better traced with \ce{HCO+} emission at inclinations close to face-on.

        To explore how spatial resolution affects the measurement of different snowlines positions, simulated line emission maps are convolved with three spatial resolutions (beams): 2$\arcsec$, 0.25$\arcsec$, and 0.1$\arcsec$ (Fig.~\ref{fig:RTbeam}).
        These spatial resolutions are representative of ALMA configurations C-1, C-5, and C-7 in Band 6, and C-2, C-6, and C-8 for Band 3, respectively.
        The distance for the simulated line emission maps is set to 293 pc (Perseus, \citealt{ortiz2018}), hence the three beams represent physical scales of $\sim$600, 70, and 30 AU, respectively.
        The peak of \ce{N2H+} emission in the case without a disk does not shift with different spatial resolutions.
        In the case with a disk, the \ce{N2H+} peak moves inward ($<$100 AU) with increasing beam size.
        Lower spatial resolutions are better suited to trace the \ce{CO} snowline location with \ce{N2H+} line emission.
        The reason is two-fold: the \ce{CO} snowline position is usually located at radii $\gtrsim$1000 AU, and higher spatial resolutions can recover the peak but not the extended emission which is necessary to properly characterize the \ce{CO} snowline location from observations.
        Based on the physico-chemical models (Section~\ref{sec:results}), the \ce{H2O} snowline location will extend out to a few 100 AU at most.
        In addition, the \ce{HCO+} gap is only marginally seen at $15^{\circ} < i \leq$ 45$^{\circ}$. 
        Thus, a spatial resolution of 2$\arcsec$ cannot resolve the \ce{H2O} snowline location with \ce{HCO+} line emission.
        A spatial resolution of 0.1$\arcsec$ would be able to unambiguously detect the \ce{HCO+} gap for a protostar with luminosity of 10 L$_{\odot}$ (the fiducial model) at the distance of Perseus.
        A beam of 0.25$\arcsec$ can also detect the \ce{HCO+} gap for $L_{\rm cen}$ $\geq$ 10 L$_{\odot}$, but would benefit from high signal-to-noise observations.
        The \ce{HCO+} gap may not be observable for sources with lower luminosities unless the spatial resolution can trace few AU scales.
        In contrast, the \ce{HCO+} gap will be easily observable with lower spatial resolutions for much more luminous sources, or those that have undergone a strong luminosity burst in the recent past.

        \subsection{\ce{N2H+} and \ce{HCO+} peak radius versus luminosity}
        \label{subsec:radvsparam}
        
\begin{figure*}
                \centering
                \includegraphics[width=\linewidth]{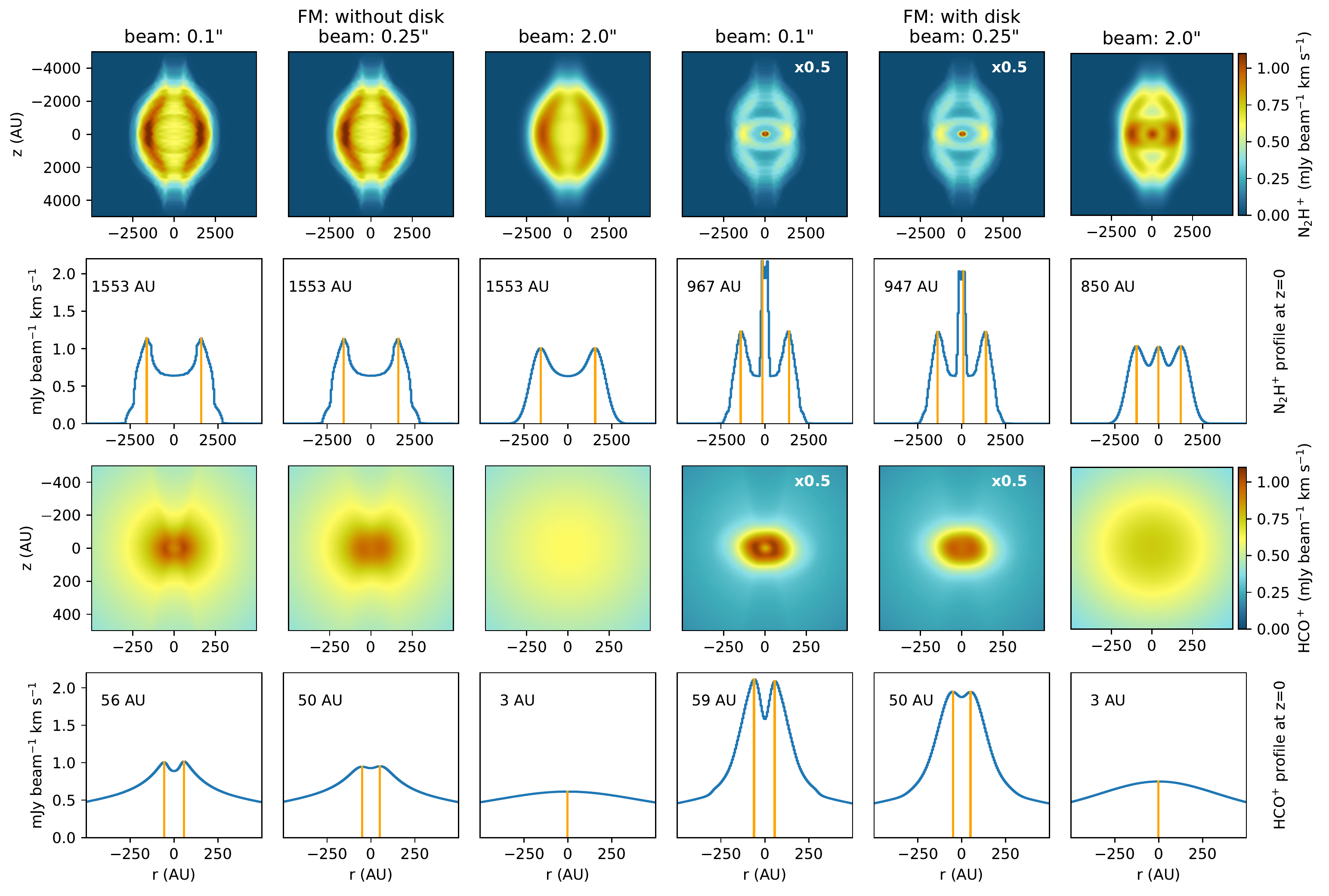}
                \caption{Simulated line emission maps showing the effect of spatial resolution (0.1$\arcsec$, 0.25$\arcsec$, and 2$\arcsec$) on the measured emission peak positions of \ce{N2H+} and \ce{HCO+}. All models have an inclination of 45$^{\circ}$. The first three columns show the fiducial model without disk. The other three columns show the fiducial model with disk. The second and third rows show corresponding slices extracted along the $z = 0$, with the number on the top left corner indicating the distance in AU of the peak from the center (i.e., peak radius). These positions are shown with orange vertical lines.}
                \label{fig:RTbeam}
        \end{figure*}

\begin{figure*}
                \centering
                \includegraphics[width=0.8\linewidth]{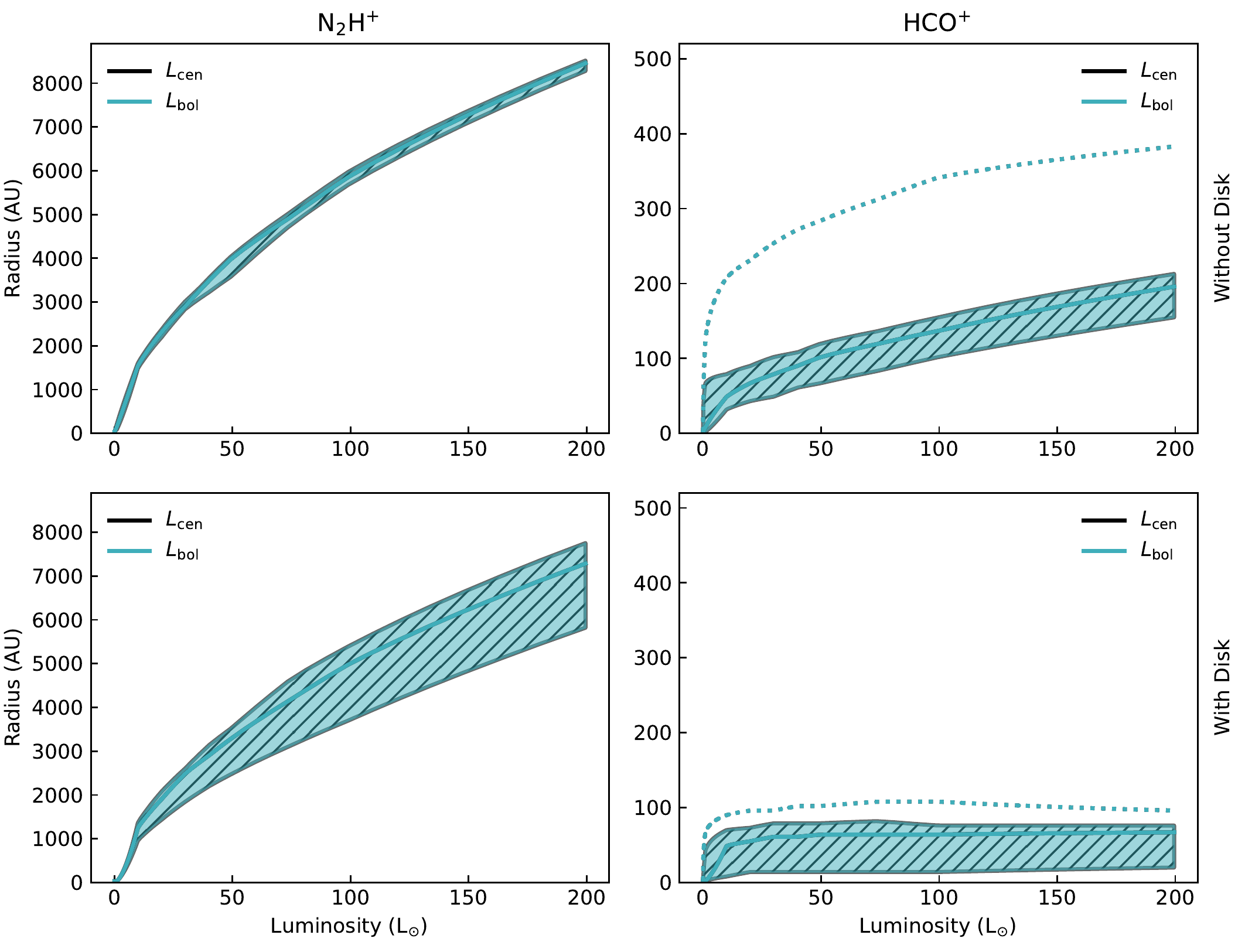}
                \caption{Peak radius of \ce{N2H+} (\textit{left column}), and \ce{HCO+} (\textit{right column}) simulated emission versus luminosity for the fiducial models without disk (\textit{top row}), and with disk (\textit{bottom row}). The black hatched area shows the central protostellar luminosity, while the colored area shows the bolometric luminosity. The solid curve shows the relation for $i$ = 45$^{\circ}$. For \ce{N2H+}, the shaded area indicates the range of peak radii for inclinations between 0$^{\circ}$ (face-on) and 90$^{\circ}$ (edge-on). For \ce{HCO+}, the shaded area indicates the range of peak radii for inclinations between 15$^{\circ}$ and 90$^{\circ}$, while the dotted line shows the peak radius vs luminosity for $i$ = 0$^{\circ}$ (face-on).}
                \label{fig:LcenLbol_vspeak}
        \end{figure*}

\begin{figure*}
                \centering
                \includegraphics[width=0.8\linewidth]{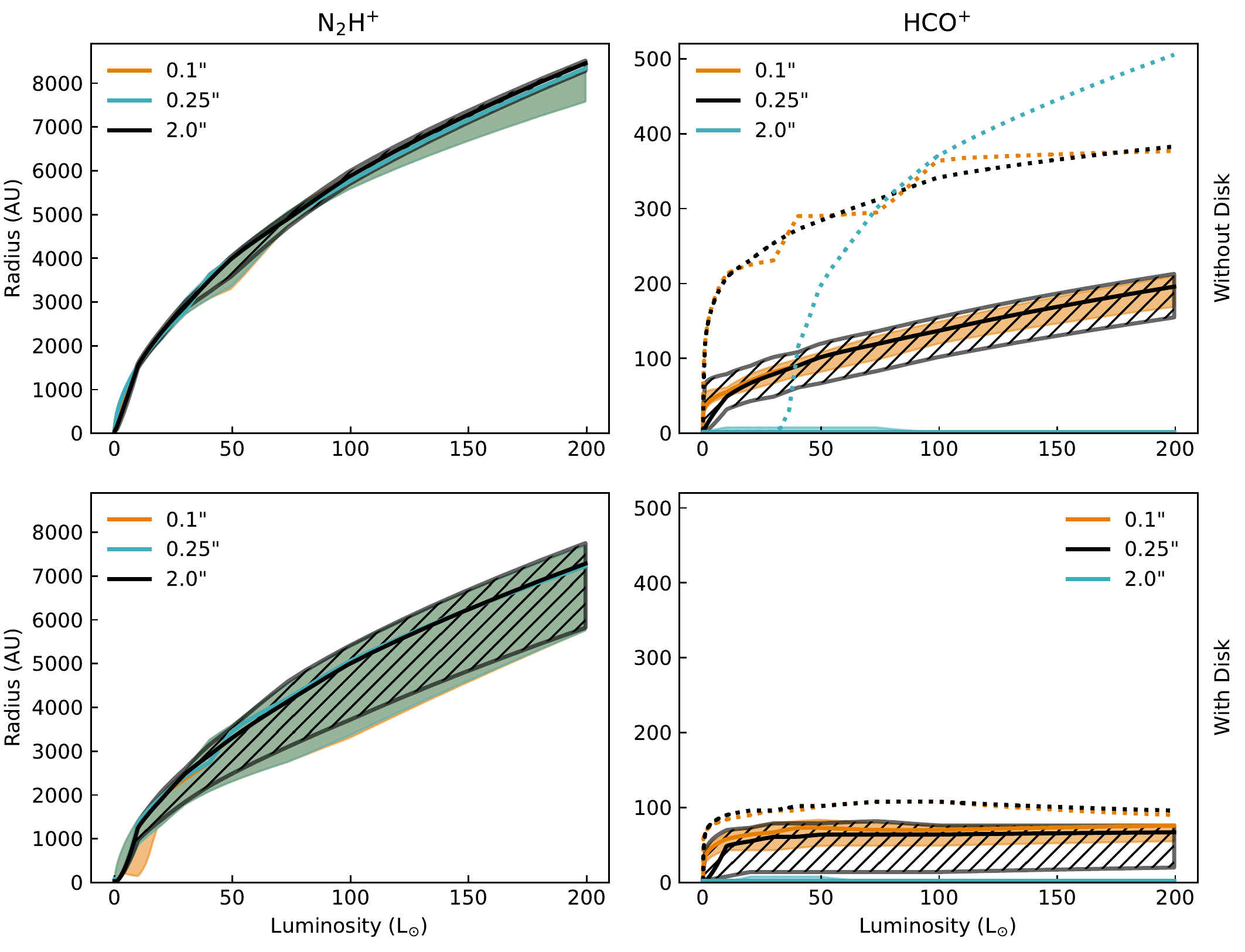}
                \caption{Peak radius of \ce{N2H+} (\textit{left column}), and \ce{HCO+} (\textit{right column}) simulated emission versus central protostellar luminosity for the fiducial models (black hatch) without disk (\textit{top row}), and with disk (\textit{bottom row}). The different shaded areas show how the measured peak radius is altered by the spatial resolution used for observations. The solid curve shows the relation for $i$ = 45$^{\circ}$. For \ce{N2H+}, the shaded area indicates the range of peak radii for inclinations between 0$^{\circ}$ (face-on) and 90$^{\circ}$ (edge-on). For \ce{HCO+}, the shaded area indicates the range of peak radii for inclinations between 15$^{\circ}$ and 90$^{\circ}$, while the dotted line shows the peak radius vs luminosity for $i$ = 0$^{\circ}$ (edge-on).}
                \label{fig:beam_vspeak}
        \end{figure*}

\begin{figure*}
                \centering
                \includegraphics[width=0.8\linewidth]{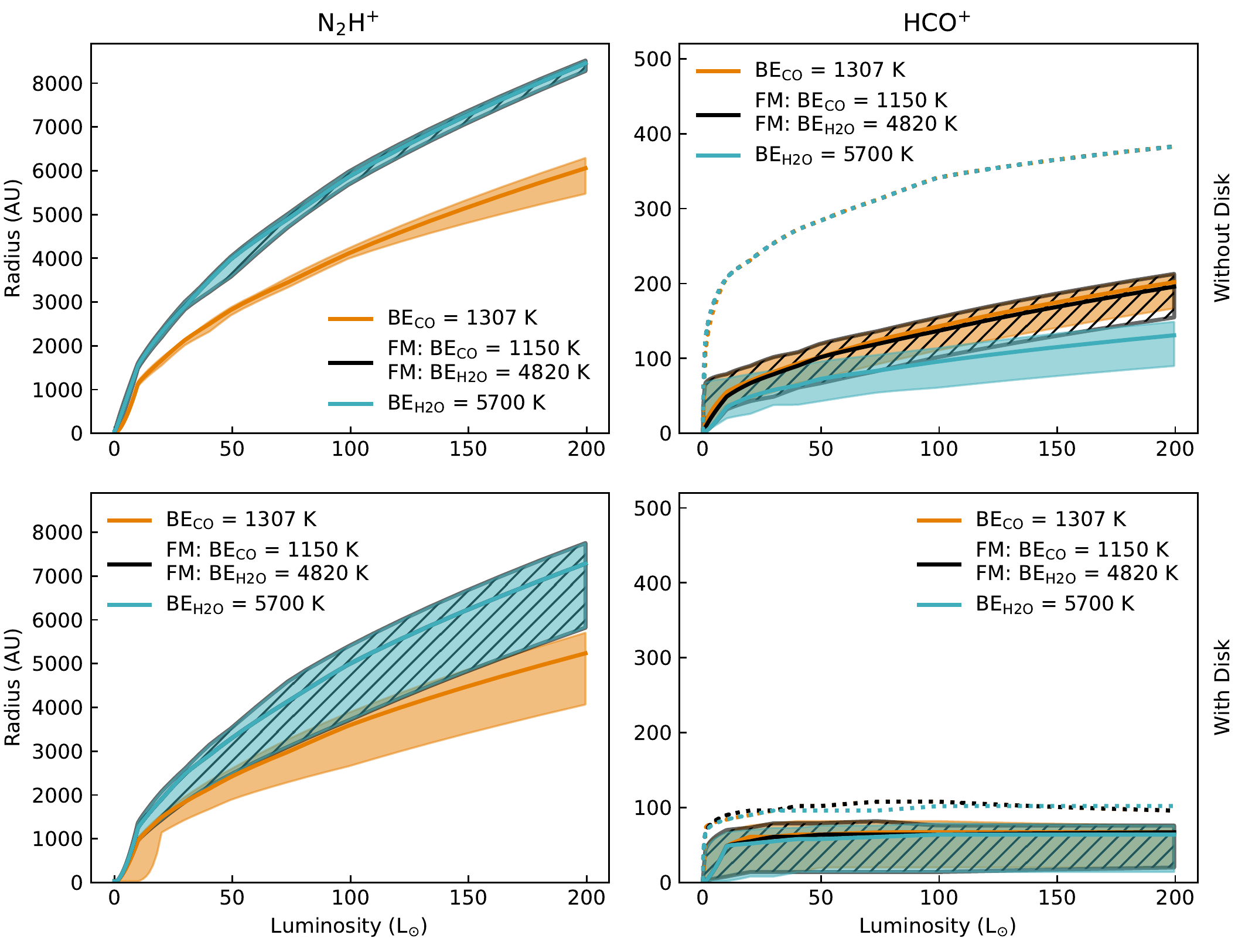}
                \caption{Peak radius of \ce{N2H+} (\textit{left column}), and \ce{HCO+} (\textit{right column}) simulated emission versus central protostellar luminosity for the fiducial models (black hatch) without disk (\textit{top row}), and with disk (\textit{bottom row}). The shaded areas show the effect of \ce{CO} and \ce{H2O} binding energy on the peak radius of \ce{N2H+} and \ce{HCO+}. The solid curve shows the relation for $i$ = 45$^{\circ}$. For \ce{N2H+}, the shaded area indicates the range of peak radii for inclinations between 0$^{\circ}$ (face-on) and 90$^{\circ}$ (edge-on). For \ce{HCO+}, the shaded area indicates the range of peak radii for inclinations between 15$^{\circ}$ and 90$^{\circ}$, while the dotted line shows the peak radius vs luminosity for $i$ = 0$^{\circ}$ (face-on).}
                \label{fig:BE_vspeak}
        \end{figure*}

\begin{figure*}
                \centering
                \includegraphics[width=0.8\linewidth]{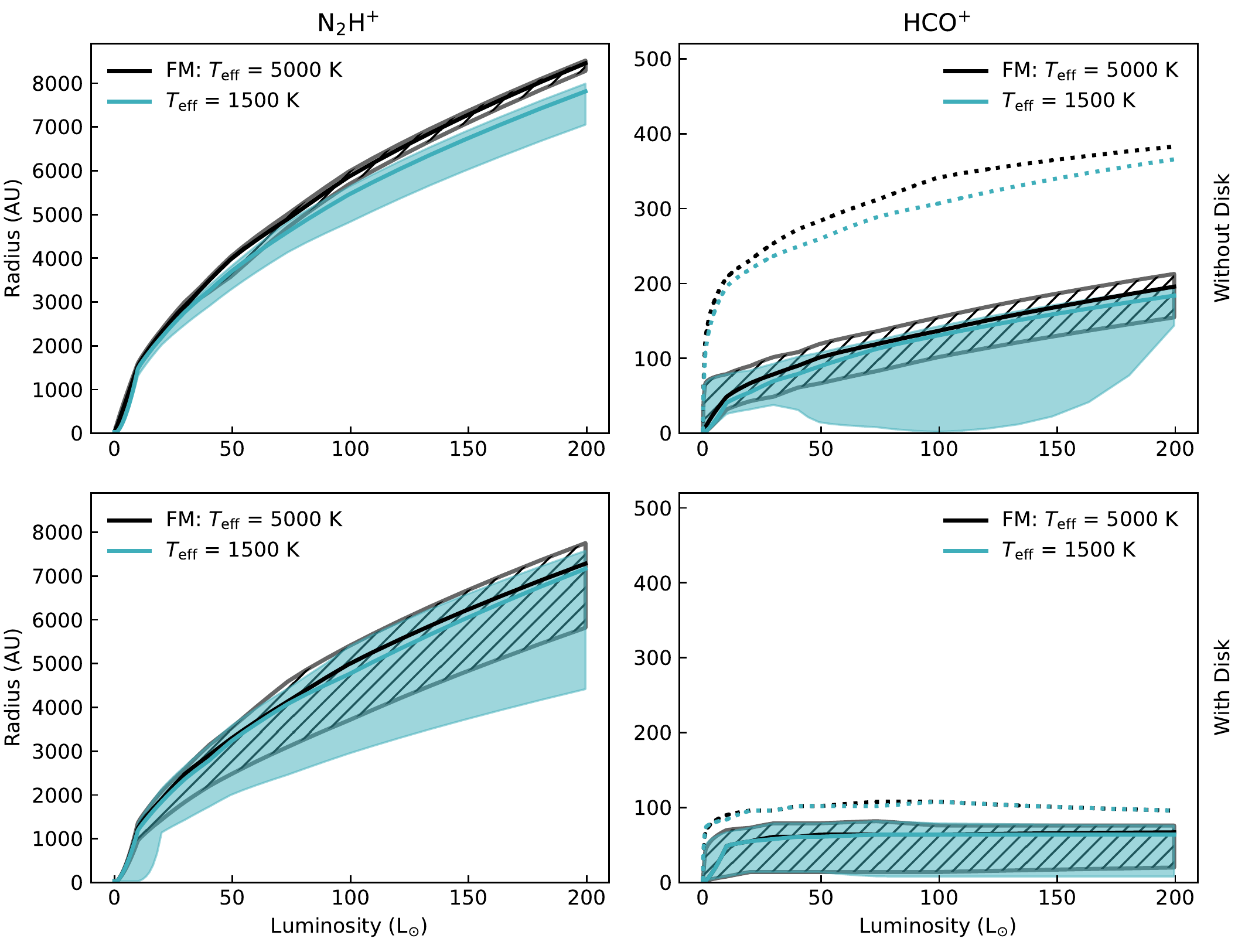}
                \caption{Peak radius of \ce{N2H+} (\textit{left column}), and \ce{HCO+} (\textit{right column}) simulated emission versus central protostellar luminosity for the fiducial models (black hatch) without disk (\textit{top row}), and with disk (\textit{bottom row}). The shaded areas show the impact of effective temperature on the peak radius of \ce{N2H+} and \ce{HCO+}. The solid curve shows the relation for $i$ = 45$^{\circ}$. For \ce{N2H+}, the shaded area indicates the range of peak radii for inclinations between 0$^{\circ}$ (face-on) and 90$^{\circ}$ (edge-on). For \ce{HCO+}, the shaded area indicates the range of peak radii for inclinations between 15$^{\circ}$ and 90$^{\circ}$, while the dotted line shows the peak radius vs luminosity for $i$ = 0$^{\circ}$ (face-on).}
                \label{fig:teff_vspeak}
        \end{figure*}

\begin{figure*}
                \centering
                \includegraphics[width=0.8\linewidth]{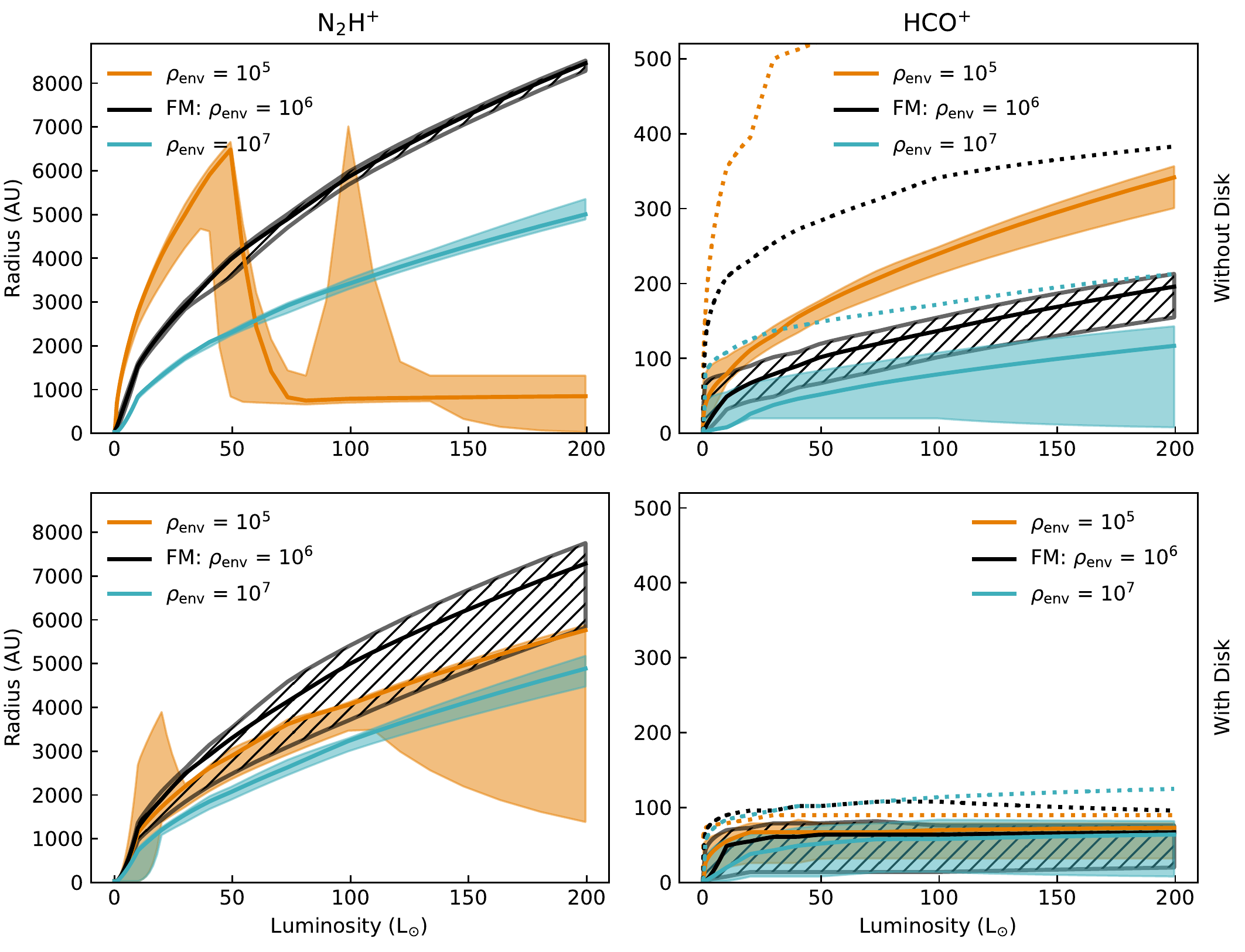}
                \caption{Peak radius of \ce{N2H+} (\textit{left column}), and \ce{HCO+} (\textit{right column}) simulated emission versus central protostellar luminosity for the fiducial models (black hatch) without disk (\textit{top row}), and with disk (\textit{bottom row}). The shaded areas show how the measured peak radius shifts with changes in envelope density. The \ce{N2H+} curve for $\rho_{\rm env}$ = 10$^{5}$ cm$^{-3}$ is not indicative of the \ce{CO} snowline due to the low density, and thus caution must be taken when comparing with observations. See main text for further discussion. The solid curve shows the relation for $i$ = 45$^{\circ}$. For \ce{N2H+}, the shaded area indicates the range of peak radii for inclinations between 0$^{\circ}$ (face-on) and 90$^{\circ}$ (edge-on). For \ce{HCO+}, the shaded area indicates the range of peak radii for inclinations between 15$^{\circ}$ and 90$^{\circ}$, while the dotted line shows the peak radius vs luminosity for $i$ = 0$^{\circ}$ (face-on).}
                \label{fig:rhoenv_vspeak}
        \end{figure*}

\begin{figure*}
                \centering
                \includegraphics[width=0.8\linewidth]{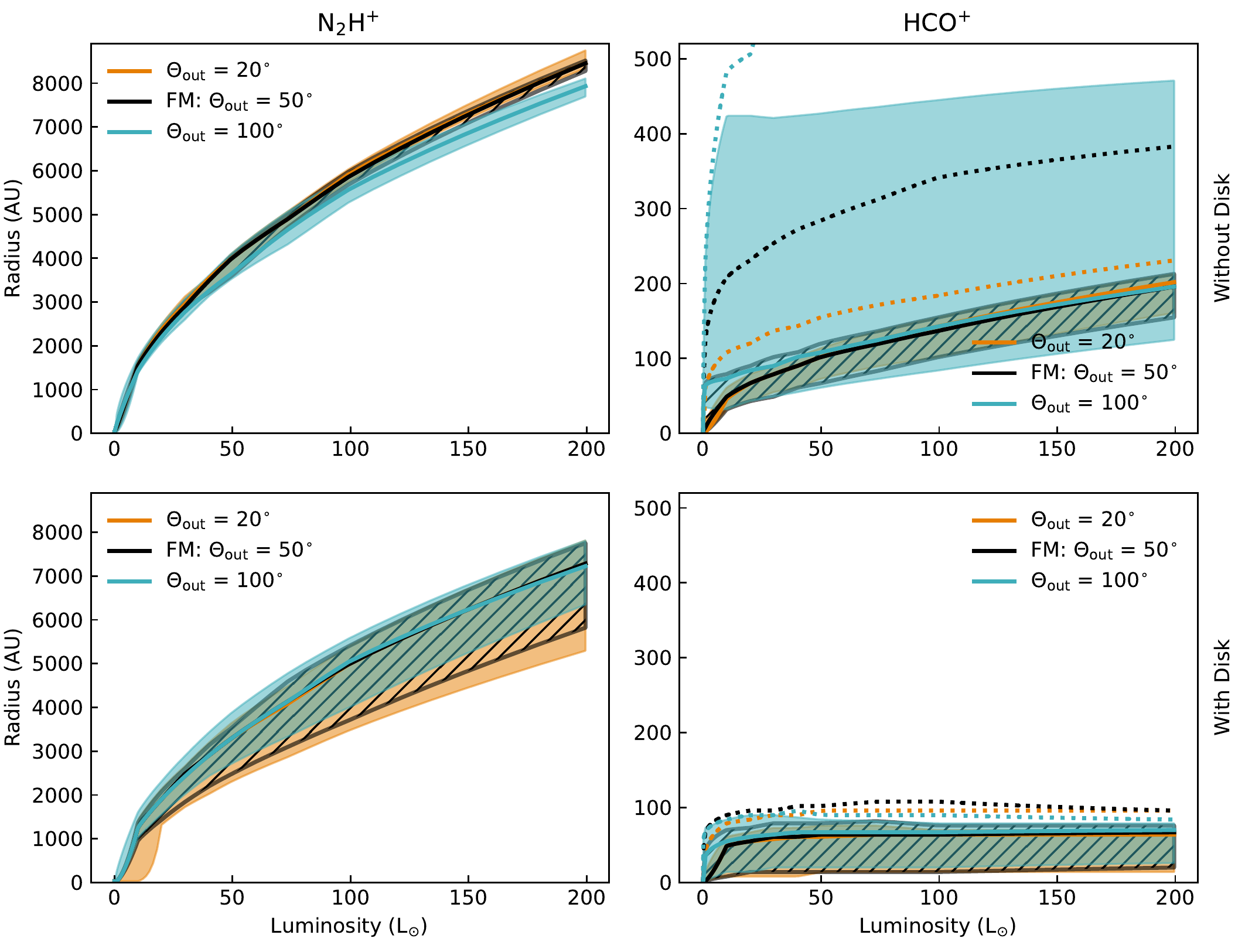}
                \caption{Peak radius of \ce{N2H+} (\textit{left column}), and \ce{HCO+} (\textit{right column}) simulated emission versus central protostellar luminosity for the fiducial models (black hatch) without disk (\textit{top row}), and with disk (\textit{bottom row}). The shaded areas show the effect of outflow cavity opening angle $\Theta_{\rm out}$ on the peak radius of \ce{N2H+} and \ce{HCO+}. The solid curve shows the relation for $i$ = 45$^{\circ}$. For \ce{N2H+}, the shaded area indicates the range of peak radii for inclinations between 0$^{\circ}$ (face-on) and 90$^{\circ}$ (edge-on). For \ce{HCO+}, the shaded area indicates the range of peak radii for inclinations between 15$^{\circ}$ and 90$^{\circ}$, while the dotted line shows the peak radius vs luminosity for $i$ = 0$^{\circ}$ (face-on).}
                \label{fig:outflow_vspeak}
        \end{figure*}

\begin{figure*}
                \centering
                \includegraphics[width=0.8\linewidth]{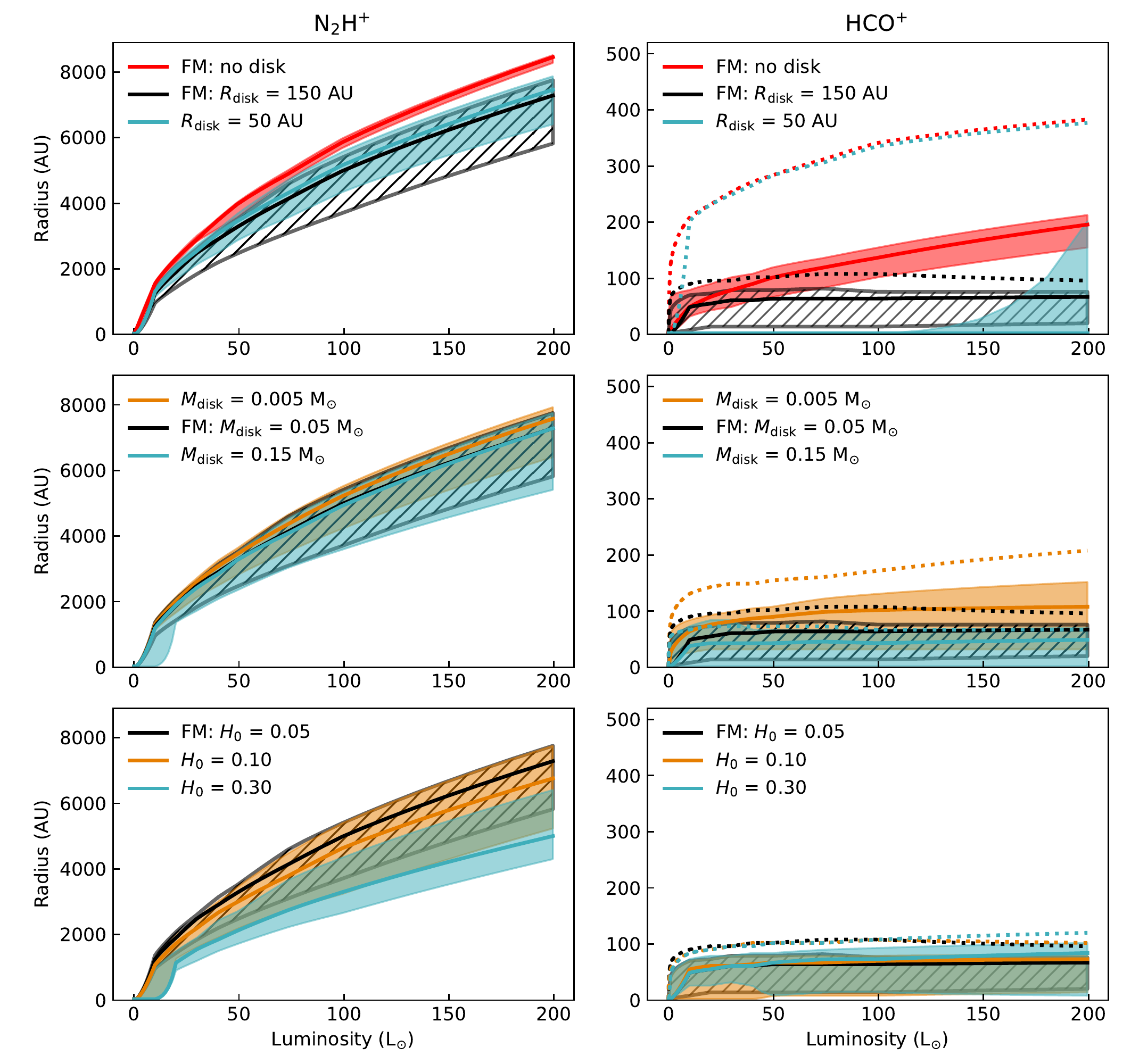}
                \caption{Peak radius of \ce{N2H+} (\textit{left column}), and \ce{HCO+} (\textit{right column}) simulated emission versus central protostellar luminosity for the fiducial models with disk (black hatch). For comparison, the top row shows the fiducial model without disk in red. The shaded areas show the effect of different disk parameters ($R_{\rm disk}$, $M_{\rm disk}$, and $H_{\rm 0}$) on the peak radius of \ce{N2H+} and \ce{HCO+}. The solid curve shows the relation for $i$ = 45$^{\circ}$. For \ce{N2H+}, the shaded area indicates the range of peak radii for inclinations between 0$^{\circ}$ (face-on) and 90$^{\circ}$ (edge-on). For \ce{HCO+}, the shaded area indicates the range of peak radii for inclinations between 15$^{\circ}$ and 90$^{\circ}$, while the dotted line shows the peak radius vs luminosity for $i$ = 0$^{\circ}$ (face-on).}
                \label{fig:disk_vspeak}
        \end{figure*}

        In order to provide a quick way to compare the models presented here with observations, plots of the \ce{N2H+} and \ce{HCO+} line emission peak radius are provided in Figures~\ref{fig:LcenLbol_vspeak} to \ref{fig:disk_vspeak}.
        Since inclination impacts the measured peak of \ce{N2H+} and \ce{HCO+}, the range of modeled inclinations are shown as shaded areas, with a curve showing the trend for $i$ = 45$^{\circ}$.
        Section~\ref{subsec:postproc} demonstrated the relevance of spatial resolution in measuring the emission peaks of \ce{N2H+} and \ce{HCO+}.
        Figure~\ref{fig:beam_vspeak} shows the three spatial resolutions (beam sizes) versus luminosity. The three spatial resolutions are described in Section~\ref{subsec:postproc} and the simulated emission maps are shown in Fig.~\ref{fig:RTbeam}.
        
        The luminosity in the plots is that of the central protostar ($L_{\rm cen}$).
        Thus, Fig.~\ref{fig:LcenLbol_vspeak} shows a comparison of $L_{\rm cen}$ with the bolometric luminosity $L_{\rm bol}$ derived from the simulated line emission maps.
        While there is a difference between $L_{\rm cen}$ and $L_{\rm bol}$, the trends are similar.
        
        The effect of low envelope density on \ce{N2H+} as a tracer of the \ce{CO} snowline location can be better visualized when comparing peak radius versus luminosity for different envelope densities (Fig.~\ref{fig:rhoenv_vspeak}).
        As noted in Section~\ref{subsec:COline}, the freeze-out of molecules at low densities is much slower, and thus the snowline locations are no longer clearly defined.
    The \ce{N2H+} curve at $\rho_{\rm env}$ = 10$^{5}$ cm$^{-3}$ in Figure~\ref{fig:rhoenv_vspeak} is not indicative of the \ce{CO} snowline location.
    The curve drops out at luminosities $>$50 L$_{\odot}$ because the \ce{N2H+} emission has become so diffuse and extended that a clear peak is no longer discernible.
    This is because the envelope density has become so low that the reaction between \ce{N2H+} and \ce{CO} (ID 13 in Table~\ref{tab:chemnetwork}) is no longer a relevant destruction path for \ce{N2H+}, hence it is present in the gas phase in a larger region of the envelope and not correlated to the distribution of \ce{CO}.
    The \ce{N2H+} curve at $\rho_{\rm env}$ = 10$^{5}$ cm$^{-3}$ in Figure~\ref{fig:rhoenv_vspeak} is kept as a reference, but caution must be taken when comparing this model with observations.
    This result illustrates that the snowline location of molecules also depends on density, and consequently so does the robustness of a molecular tracer for a particular snowline location. 
        
        An interesting result worth noting is how the presence of a disk limits the outward shift of the \ce{H2O} snowline position.
        This effect is somewhat dependent on luminosity and density.
        For luminosities $<$ 10 L$_{\odot}$, the \ce{H2O} snowline position increases sharply up to about 50 AU, regardless of which parameter is changed.
        For luminosities $\geq$ 10 L$_{\odot}$, the water snowline position is limited to within radii below 150 AU.
        The \ce{H2O} snowline position moves out to larger radii with decreasing disk density.
        Thus, for a disk with the same radius and scale height, a lower disk mass will allow the snowline position to move out further than a higher disk mass (Fig.~\ref{fig:density})
        In contrast, when a disk is present the \ce{CO} snowline location shifts inward by $\sim$1000 AU along the disk mid-plane relative to the models without a disk.
        However, the \ce{CO} snowline location still shifts to larger radii with increasing luminosity when a disk is present.

        \section{Discussion}
        \label{sec:discussion}
        
        \subsection{Caveats}
        \label{subsec:caveats}
        
        Several caveats to the models presented in this work should be noted.
        The chemical network only takes into account a limited set of simple molecules (three atoms or less per molecule) and a reduced number of reactions.
        The reduced chemical network in this work does not allow the additional chemical processes to be studied in the context of the \ce{CO} and \ce{H2O} snowline locations. 
        Examples of such processes are alternative formation and destruction paths for molecular species in the network (e.g., warm \ce{DCO+}, \citealt{favre2015,murillo2018a}), extent of other simple molecules (e.g., \ce{SO}, \ce{CN}), and the relation between complex molecular species and snowline position.
        Since water is given a simple treatment in our network, additional effects on the \ce{H2O} snowline location from water chemistry cannot be characterized or discussed here.
        It is known from previous observational studies with \textit{Herschel} (e.g., \citealt{kristensen2012}) that gas-phase water is also found along the outflow cavity.
        We do not include processes such as photodissociation and photoionization in our models, and so the chemistry in the outflow cavity is not accurately modeled.
        Hence, the calculated abundances within the outflow cavity are not shown in the molecular distributions and simulated line emission maps.
        This could alter the distribution and measured peak radius of the \ce{HCO+} emission, and consequently the inferred water snowline location.
        It is also possible that wider cavities could lead to more of the envelope being heated, leading to more water in the envelope, and a different distribution of \ce{HCO+} emission.
        To accurately simulate the chemical composition of outflow cavity walls, photodissociation and hot gas-phase chemistry are needed (e.g., \citealt{drozdovskaya2015}).
        
        Opacity also plays a role in the simulated line emission maps and the measured emission peak location, especially for \ce{HCO+}.
        In our models, optical depths are mainly determined by the density and kinematic structure.
        If the line emission is optically thick, the intensity map cannot properly reflect the molecular spatial distribution (e.g., \ce{HCO+}, \citealt{vanthoff2021}).
        It is also worth noting that dust grain growth can occur in the early evolutionary stages and can block line emission from the region immediately around a protostar (e.g., \citealt{harsono2018}).
        This could result in a ring structure after continuum subtraction \citep{lee2019}.
        No continuum subtraction is performed on our simulated line emission images to avoid this artificial chemical substructure.
        
        \subsection{Robustness of \ce{N2H+} and \ce{HCO+} as snowline position tracers}
        \label{subsec:diss_robust}
        
        Observational studies of \ce{CO} and \ce{H2O} snowline positions typically use the peak emission of \ce{N2H+} and \ce{HCO+} to trace the respective snowline locations (e.g., \citealt{qi2015,hsieh2018,vanthoff2018a,hsieh2019b,qi2019}). 
        The simulated line emission maps discussed in Section~\ref{sec:results} examined the robustness of \ce{N2H+} and \ce{HCO+} as snowline position tracers, as well as \ce{DCO+}, within the context of the limited chemical network used here.
        To fully assess whether \ce{N2H+} and \ce{DCO+} are robust tracers of the \ce{CO} snowline location, and \ce{HCO+} of the \ce{H2O} snowline location, a complete chemical network which includes all formation and destruction pathways for the relevant species is needed. The current work aims to find the conditions that produce significant and observable impacts on the \ce{CO} and \ce{H2O} snowline locations. Thus, complexity is sacrificed for computational speed. The use of a more complete network informed by the results in this work will be investigated in the future.
        
        Previous studies of the robustness of \ce{N2H+} as a tracer of the \ce{CO} snowline location focused on isolated protoplanetary disks (few 100 AU; \citealt{vanthoff2017}).
        This work examines the robustness of \ce{N2H+} in the envelope (few 1000 AU) during the early embedded protostellar phase.
        The robustness of \ce{HCO+} was studied observationally in \citet{vanthoff2018a}, and was argued to be a good tracer of the \ce{H2O} snowline position.
        However, the optical depth of the \ce{HCO+} emission can affect the real peak radius related to the snowline location \citep{vanthoff2020}.
        In our results, we find that by changing the optical depth (i.e., $\rho_{\rm env}$), the \ce{HCO+} emission might suggest that the snowline is located at a different latitude if there is no disk present.
        We note that this process is also affected by the velocity structure and inclination which determines the optical depth at a specific velocity range.
        A possible solution might be to observe a range of isotopologues with different abundances that are less optically thick (e.g., \citealt{vanthoff2021}).
        
        Although the peak of the \ce{N2H+} emission does not trace the exact position of the \ce{CO} snowline, it does provide an accessible observational proxy, except at low envelope densities where \ce{N2H+} no longer traces the \ce{CO} snowline location (see Sect.~\ref{subsec:radvsparam} and Fig.~\ref{fig:rhoenv_vspeak}).
        \citet{vanthoff2017} also found that \ce{N2H+} peaks further out than the \ce{CO} snowline position in protoplanetary disks, which is consistent with the results presented here.
        While \citet{vanthoff2017} suggests chemical modeling is necessary to derive a robust location of the \ce{CO} snowline using \ce{N2H+} observations, the results could still be degenerate, and require more observational constraints.
        Constraining additional parameters, such as the envelope density, and the presence of a disk would prove to be helpful, but it would require multiwavelength observations with a range of spatial resolutions and  sufficient spectral resolution to probe the kinematics.
        Additional molecular tracers would also provide constraints regarding the \ce{CO} snowline location.
        By itself, \ce{DCO+} is not a robust tracer of the \ce{CO} snowline location since it will also strongly depend on density and chemistry (e.g., \citealt{qi2015}).
        \ce{DCO+} also has a second, warm gas formation pathway \citep{favre2015} not included in the chemical network used here.
        This warm formation pathway can enhance \ce{DCO+} in the disk region, and can be easily observed with the \ce{DCO+} 5--4 transition \citep{murillo2018a}.
        However, combining \ce{N2H+} and \ce{DCO+} observations can provide a better estimate of the \ce{CO} snowline position (Fig.~\ref{fig:COsnowlineRT}).
        Furthermore, having a second \ce{CO} snowline location tracer can prove useful in protostellar systems where the outer envelope is so cold that \ce{N2H+} (or \ce{N2D+}) is significantly less abundant and difficult to detect because of the depletion of gas-phase \ce{N2} onto dust grains (e.g., VLA 1623-2417; \citealt{murillo2015}).
        
        Similar to \ce{N2H+}, \ce{HCO+} emission peaks some distance away from the \ce{H2O} snowline location.
        However, \ce{HCO+} can still be a good observational proxy and provide an outer limit to the \ce{H2O} snowline location, as suggested by observational studies \citep{hsieh2019b,vanthoff2018a}, as long as \ce{HCO+} emission can be spatially resolved.
        Given that \ce{H2O} is in the gas phase, where the gas temperature is $\geq$100 K, other molecular species that sublimate at similar temperatures, such as methanol, could also be used as independent tracers of the approximate location of the \ce{H2O} snowline.
        However, similar to \ce{HCO+}, their successful detection could be dependent on source inclination, spatial resolution, and presence of a disk.
        
        \subsection{Chemical effects}
        \label{subsec:diss_chem}
        
        The binding energy of molecules onto dust grains depends on the conditions under which the molecule sticks to the grain.
        Binding energies of molecules in ``pure'' ices are typically lower than when the same species are in mixtures, for example when mixed with water ice (e.g., \citealt{martindomenech2014})
        This has an impact on the sublimation temperature of a particular molecule, which in turn can alter its snowline location.
        Both molecular distributions and simulated line emission maps reflect this behavior (Fig.~\ref{fig:BE} and \ref{fig:BEH2O}), showing that the \ce{CO} and \ce{H2O} snowline locations move inward when their respective binding energies are increased.
        This effect is further highlighted when considering how the peak radius of \ce{N2H+} and \ce{HCO+} emission varies with luminosity (Fig.~\ref{fig:BE_vspeak}).

        \subsection{Heating sources}
        \label{subsec:dis_heating}
        
        The effective temperature $T_{\rm eff}$ of the protostar has a minor effect on the positions of the \ce{CO} and \ce{H2O} snowlines (Fig.~\ref{fig:teff_vspeak}).
        In fact, the most notable effect of $T_{\rm eff}$ is along the outflow cavity.
        For the same bolometric luminosity, the models with $T_{\rm eff}$ =  5000 K heat their outflow cavity to 50 K out to a distance 1.5 times that of the models with $T_{\rm eff}$ =  1500 K.
        This is because the central source with a higher $T_{\rm eff}$ produces more photons at short wavelengths, heating up the surrounding material more efficiently.
        This is consistent with observations that show that the majority of heating from the protostar escapes through the outflow cavity \citep{vankempen2009,yildiz2015,murillo2018a,murillo2018b}.    
        
        The protostellar luminosity $L_{\rm star}$ has a more significant effect on the temperature of the cloud core than $T_{\rm eff}$.
        Because of this, the protostellar luminosity has a much larger role in determining the positions of the \ce{CO} and \ce{H2O} snowlines.
        Comparing the luminosity of a protostar with the location of the \ce{N2H+} and \ce{HCO+} emission peaks is commonly used, in both observations and models, to determine whether a protostar has undergone an accretion burst or not \citep{harsono2015,visser2015,vanthoff2017,frimann2017,hsieh2018,hsieh2019b,jorgensen2020}.
        The models presented here are indeed consistent with this practice.
        While luminosity is relevant in setting the location of snowlines, the models presented here suggest that the cloud core conditions need to be constrained in more detail before determining if the snowline location is due solely to the luminosity of the protostar.
        Figures~\ref{fig:BE_vspeak} to \ref{fig:disk_vspeak} show that the presence or lack of a disk, and parameters such as envelope density and binding energy, can significantly change the snowline position even if the luminosity is the same.
        
        \subsection{Cloud core structure}
        \label{subsec:diss_core}
        
        The envelope density $\rho_{\rm env}$ is a key parameter in setting the \ce{CO} and \ce{H2O} snowline locations (Fig.~\ref{fig:density} and \ref{fig:densityH2O}).
        Physically, the density structures regulate the radiation field by redirecting the photons (Fig.~\ref{fig:density}, top row), resulting in distinguishable thermal structures in each model.
        Changes in density are, in a way, also a chemical effect for two reasons.
        First, the fractional abundances of all molecules in the chemical network are with respect to the \ce{H2} number density.
        As input, the \ce{H2} number density is set equal to the density profile of the physical model.
        Thus changing the density profile changes the absolute abundances.
        Second, the accretion rates of \ce{CO} and \ce{H2O} are a stronger function of density ($\propto n^2$) than the thermal desorption rates. 
        Hence, the density, as well as the temperature, sets the snowline location. 
        The lower the density, the longer the freezeout times. 
        At sufficiently low densities, the freezeout timescale will be so long that the molecule will stay in the gas phase regardless of the temperature. 
        Consequently, the location of the \ce{CO} and \ce{H2O} snowlines will change with the cloud core density (Fig.~\ref{fig:density} and ~\ref{fig:densityH2O}), and snowline positions do not occur at a single, well defined temperature as commonly assumed.
        The freeze-out timescale is commonly used as a chemical clock to estimate the burst interval \citep{frimann2017,hsieh2018,hsieh2019b}.
        In such cases, the cloud core density proportional to the desorption rate would need to be constrained in order to use the freeze-out timescale as a reliable clock.
        Comparing the peak emission radius of \ce{N2H+} and \ce{HCO+} with luminosity when considering different envelope densities (Fig.~\ref{fig:rhoenv_vspeak}) shows that for the same luminosity, the envelope density can drastically alter the peak emission radius, unless there is a disk present.
        
        As the protostar evolves, the envelope is expected to disperse either due to accretion onto the protostar or by material being pushed out by the outflow.
        In turn the outflow cavity is expected to widen as the protostar evolves \citep{arce2006}. 
        Because heating from the protostar mainly escapes through the outflow cavity, wider outflow cavities allow more of the envelope to be heated by the protostar.
        Despite this, the emission peaks of \ce{N2H+} and \ce{HCO+} in our model present no significant shift with increasing outflow cavity opening angle (Fig.~\ref{fig:outflow_vspeak}).
        Thus, the chemical model results and simulated line emission maps do not suggest that the outflow cavity angle has a strong impact on the location of the \ce{CO} and \ce{H2O} snowlines.
        However, outflow cavity wall chemistry needs to be added to accurately study how the outflow cavity opening angle affects the water snowline position.
        
        Observational studies have shown that dust grain growth can occur in embedded protostellar sources (e.g., \citealt{harsono2018}).
        The effect of different dust grain size distributions, dust growth, and dynamical processes of dust are not treated in this work.
        There are, however, previous papers which have studied the effect of dust grain size on snowline locations.
        For example, \citet{panicmin2017} have studied the effect of gas and dust evolution in the disk mid-plane around typical Herbig Ae stars, while \citet{gavino2021} have studied the effect of dust grain size distribution, grain-size dependent temperature and dust settling on the gas-grain chemistry in the context of a representative T Tauri disk.
        The \citet{panicmin2017} models explore minimum dust grain sizes between 0.01 to 100 $\mu$m, and maximum dust grain sizes in the range of  1 mm to 1 km, with a power-law index of 3.5. The modeled region spans $r$ = 0.24 -- 500 AU for an A-type star with 2 M$_{\odot}$, $L_{star}$ = 35 L$_{\odot}$ and $T_{eff}$ = 10~000 K.
        The \citet{gavino2021} models present a range of dust grains between 5 nm to 1 mm, with a power-law index of 3.5 within a region up to $r$ = 30 -- 250 AU and $z/r$ = 0 -- 0.5, for a pre-main sequence star with $L_{star}$ = 0.75 L$_{\odot}$ and $T_{eff}$ = 3900 K.
        Meanwhile, our models of the cloud core of embedded protostars ($r, z$ = 8790 AU) include a grain-size distribution of 0.5 -- 100 $\mu$m for the disk, and 0.5 -- 1 $\mu$m for the envelope, both with a power-law index of 3.5, $L_{star}$ = 0.01 -- 200 L$_{\odot}$ and $T_{eff}$ = 1500 or 5000 K.
        Despite the differences, the key results of all model grids can be compared.
        
        The lack of grains $\leq$ 0.01 $\mu$m in our models avoids wide temperature fluctuations such as those found in \citet{gavino2021} which would affect the \ce{CO} and \ce{H2O} snowline locations. 
        \citet{panicmin2017} note that the temperature profile in the disk depends on both the total gas mass and the size of the smallest grains. In particular, dust growth leads to a decreased temperature given dust settling and reduced $\tau$ = 1 surfaces in the gas.
        Our models confirm that the effect of reducing the envelope density parameter, and thus the total gas mass, leads to lower temperatures in disk, in particular along the mid-plane, even when the dust grain size does not change (Fig.~\ref{fig:densityH2O}). In contrast, the envelope shows an increase in temperature when the total gas mass is reduced (Figs.~\ref{fig:density} \& \ref{fig:densityH2O}). The different effects in the disk mid-plane and envelope are most likely product of the dust grain sizes in the disk (0.5 -- 100 $\mu$m) and envelope (0.5 -- 1 $\mu$m).
        
        While we do not consider the vertical water snowline location, and \citet{gavino2021} cannot provide constraints on the radial water snowline location ($<<$ 30 AU), both models agree on two aspects regarding the water snowline. 
        The \ce{H2O} snowline location is regulated by the penetration of UV and the amount of \ce{H2} in the disk; and water in the gas phase is mainly present in the warm regions (upper layers) of the disk. 
        These aspects are shown in our model by the fact that the presence of the disk, and its characteristics, limits the water snowline location, and the gas-phase water is located between the outflow cavity wall and the disk surface. 
        When a disk is not present, and thus the largest grains are 1 $\mu$m, the water snowline location shifts outward with increasing luminosity.
        \citet{gavino2021} points out that dust temperature impacts chemistry. 
        Our models demonstrate that the presence of a disk, with larger dust grains, significantly changes the temperature profile of the cloud core, and consequently the snowline location of different molecular species. 
        We further highlight that gas density also plays a key role in the distribution of different molecular species.
        
        Given that this work models embedded protostars with significant envelope, while \citet{panicmin2017} and \citet{gavino2021} model disks around Ae Herbig and T Tauri sources, the \ce{CO} snowline location cannot be directly compared. 
        However, the UV shielding and grain temperature relation to freeze-out effects for \ce{CO} reported by \citet{gavino2021} are seen in our models as well. Similarly, the importance of gas mass and smallest grain sizes on the \ce{CO} snowline location found in \citet{panicmin2017} is also reproduced.
        Changes in envelope gas mass alter the \ce{CO} snowline, whether a disk is present or not.
        The presence of a disk, which has a larger range of dust grain sizes, changes the \ce{CO} snowline location and distribution of \ce{CO} in the gas phase throughout the cloud core relative to the case without a disk.
        Indeed, a flared disk further produces a change in the \ce{CO} gas phase distribution at the edge of the disk along the mid-plane (Fig.~\ref{fig:COdiskmodel}).
        In observational studies of embedded protostars this effect is often referred to as disk shadowing (e.g., \citealt{murillo2015}), which is basically UV shielding of the envelope by the dust disk causing change in the dust temperature in the envelope along the disk plane.
    While this comparison provides some insight into the effect of dust grain size distributions on molecular gas, it does not provide the full picture for the case of embedded protostars and snowline locations.
    For example, it is not clear how different ratios of large to small dust grains within the envelope and disk (i.e., the power-law index), and a wider range of dust grain sizes affect the molecular gas distribution.
        This topic is left for future work.

        \subsection{Effect of disk geometry}
        \label{subsec:diss_diskgeo}
        
        In all the molecular distribution and simulated line emission maps the presence of a disk generates a significant change in the temperature structure of the cloud core.
        When a disk is present, both \ce{CO} and \ce{H2O} snowline positions are located at smaller radii relative to the case without a disk (Fig.~\ref{fig:disk_vspeak}).
        The presence of the disk causes \ce{N2H+} emission peak to move inward along the disk mid-plane, producing an hour-glass-like morphology.
        In contrast, the extent of the \ce{HCO+} emission becomes more compact when a disk is present (Figs.~\ref{fig:H2Odiskmodel} and \ref{fig:H2OsnowlineRT}).
        
        The presence of a disk (Fig.~\ref{fig:fiducial}), its radius, mass, scale height and density (Figs.~\ref{fig:COdiskmodel}, \ref{fig:H2Odiskmodel}, \ref{fig:disk_vspeak}) alter the temperature and chemical structure throughout the core.
        The molecular distributions alone would suggest that the disk mainly affects regions beyond the disk edge.
        However, the simulated line emission maps show that even the inner regions are affected by the presence of a disk, no matter how small.
        
        For \ce{HCO+} and \ce{H2O} emission, the presence of a disk regulates the radial extent of their emission and thus the \ce{H2O} snowline location (Figs.~\ref{fig:H2OsnowlineRT} and \ref{fig:disk_vspeak}).
        These results suggest that the presence of a disk in embedded cloud cores could help limit the radial extent of warm molecules in protostellar cloud cores to regions within the disk.
        This of course is true if the only heating source is the central protostar.
        An external heating source, such as accretion shocks or another protostar, may change the distribution of warm molecules through the disk.
        The first row of Fig.~\ref{fig:disk_vspeak} shows that a disk of $R_{\rm disk}$ = 50 AU and $M_{\rm disk}$ = 0.05 M$_{\odot}$ causes the \ce{HCO+} emission to be very compact, unless the protostar reaches luminosities of 100 L$_{\odot}$ or higher.
        This could provide an argument as to why very bright sources, or massive protostars, appear so chemically rich out to large radii in comparison with low mass protostellar sources even when a disk is present.
        Given the shape of disks, inclination plays a role in whether regions within the \ce{H2O} snowline position can be observed.
        However, we cannot necessarily know the inclination of a disk in advance, especially if it is small, below the spatial resolution of the data, or there is no kinematic data to infer the presence of a disk.
        If emission from warm molecules is detected then the disk, if present, is close to face-on ($i < 45^{\circ}$) and the snowline location can be measured.
        On the other hand, if warm molecules are not observed (despite other indicators that they should be, e.g., bolometric luminosity), then it might suggest that a disk, if present, is close to edge-on ($i > 45^{\circ}$), and measurement of the \ce{H2O} snowline location becomes more challenging. 
        
        Thus, the disk behaves like an ``umbrella.'' 
        The disk shields the cloud core from protostellar heating along the mid-plane, while at the same time it retains the warm (and hot) gas at scales within the disk radius.
        The presence of disks in embedded sources, both rotationally supported and flattened dust structures, with a range of radii and masses has been confirmed observationally 
        \citep{jorgensen2009,enoch2011,murillo2013,harsono2014,yen2015,persson2016,yen2017,maret2020,tobin2020}.
        In addition, molecular species tracing warm regions have a tendency to be observed in the inner regions of the protostar and along the disk-like structures or the outflow cavity (e.g., \citealt{murillo2018a,artur2019}).
        Further observational evidence of the disk altering the location of cold molecules in embedded sources has also been reported \citep{murillo2015}.
        
        The presence of a disk dictates where warm and hot gas are located inside the cloud core.
        Thus, species such as water and complex organic molecules (COMs) released from the dust grains are very likely present in every protostellar source.
        Their successful detection is dependent on the presence of a disk, in combination with the protostellar luminosity, inclination angle, and spatial resolution.
        When a disk is present, water and COMs are most likely to be detected at inclinations $i \leq 45^{\circ}$.
        The necessary spatial resolution will depend on the luminosity of the protostellar system, with the lowest luminosity objects requiring spatial resolutions equivalent to $\sim$10 AU.
        At $i \sim 45^{\circ}$, the degree of disk flaring will affect how well the warm molecular regions can be detected.
        The impact of a disk on the chemistry of a cloud core could provide some insight into observations of protostellar systems, a few examples are noted here.
        The concentration of emission from warm molecules to a small region around the protostars (e.g., IRAS 16293-2422, \citealt{jorgensen2016,murillo2021}). 
        The presence of a disk, and its inclination, can obscure the warm molecular regions (e.g., NGC1333 IRAS4A, \citealt{desimone2020}).
        An aspect that would still need to be explored is how warm disks, like those in Taurus (e.g., \citealt{vanthoff2020}), affect the snowline location.
        
        If the disk does indeed help constrain the radial extent of all warm molecular species, this would have an implication for planet formation.
        Recent studies show that embedded disks have the dust mass needed to produce planets \citep{tychoniec2018,tychoniec2020}, and possibly the dust size as well \citep{harsono2018}.
        Hence, the effect seen in the results in this work would suggest that embedded disks not only produce planets, but it is the physical conditions in the embedded phase that set their chemical composition.

        \section{Conclusions}
        \label{sec:conclusions}
        This paper presents a grid of cylindrical symmetric (2D) steady-state physico-chemical models aiming to study the conditions and source parameters that affect the \ce{CO} and \ce{H2O} snowline locations within protostellar cloud cores.
        The chemical network included deuterated species and the most important ion-molecule reactions for prescribing the abundances of \ce{N2H+}, \ce{HCO+}, and \ce{DCO+}. 
        A simplified treatment of water, which does not consider the formation of \ce{H2O} but only its freeze-out and desorption from dust grains, was used.
        A range of molecular binding energies for \ce{CO} and \ce{H2O} were used to simulate different ice environments (pure ices versus mixed ices).
        For physical parameters, the effective temperature of the protostar, luminosity, cloud core density, outflow cavity opening angle, and disk geometry (radius, mass, scale height, and density) were considered.
        Two fiducial models, one with a disk and one without, were used as references to understand how each individual parameter affected the \ce{CO} and \ce{H2O} snowline positions.
        We simulated the molecular line emission from snowline location tracers with the purpose of determining which parameters produce observable effects on the snowline locations.
        With the simulated line emission maps, the robustness of \ce{N2H+} and \ce{HCO+} as snowline position tracers, the impact of inclination, and the spatial resolution on the emission peak positions as measured from the simulated line emission maps were also addressed.
        Finally, for the purpose of comparing the models with observations, plots of the peak emission radius versus luminosity for the studied parameter space are presented.
        
        The results of this study are as follows:
        \begin{enumerate}
                \item The \ce{CO} and \ce{H2O} snowline locations are mainly dictated by luminosity and cloud core density. 
                Increasing luminosity shifts the snowline location outward to larger radii.
                In contrast, increasing the protostellar cloud core density causes the snowline location to shift inward to smaller radii, regardless of the protostellar luminosity.
                \item The \ce{CO} snowline location shifts radially outward or inward in all directions when there is no disk present.
                When a disk is present, the \ce{CO} snowline position shifts inward along the disk mid-plane. 
                Vertical shifting of the \ce{CO} snowline position also occurs if the disk is flared.
                \item When no disk is present, the \ce{H2O} snowline position shifts radially outward as luminosity increases. 
                In contrast, when a disk is present, the radial shift of the \ce{H2O} snowline position along the disk mid-plane is limited to radii below the disk radius, concentrating the gas-phase water to small regions around the protostar.
                The exception to this trend occurs with small disks and high luminosities ($R_{\rm disk}$ = 50 AU and $L_{\rm star} >$ 150 L$_{\odot}$ in this work).
                This effect would also concentrate all warm (and hot) gas-phase molecules in the cloud core to small regions around the protostar.
                \item Both inclination of the protostellar cloud core along the line of sight and the spatial resolution of the data affect the observability and measurement of snowline locations, in particular when disks are present.
                The \ce{H2O} snowline location is easier to measure with inclinations $i \lesssim 45^{\circ}$ and spatial resolutions that can resolve $\sim$10 AU scales.
                At higher inclinations $i > 45^{\circ}$ and lower spatial resolutions, the \ce{H2O} snowline location is difficult to measure from \ce{HCO+} emission.
                At inclinations $i \sim 45^{\circ}$, the degree of disk flaring affects how well the \ce{H2O} snowline location can be measured.
                Measurement of the \ce{CO} snowline location is not significantly affected by inclination or spatial resolution.
                \item Finally, \ce{N2H+} and \ce{HCO+} emission serve as good observational tracers of the \ce{CO} and \ce{H2O} snowline locations.
                Determining whether a disk is present, additional molecular tracers (e.g., \ce{DCO+} and line emission from warm or hot molecules, such as methanol), and envelope density (even an upper limit) would provide additional constraints that would help in accurately determining the factors that influence the observed \ce{CO} and \ce{H2O} snowline positions.
         \end{enumerate}
         The models presented in this work show that the snowline position is not related to a single physical parameter, such as luminosity, but rather the snowline location is dependent on several factors. The physical structure of the protostellar cloud core plays a key role in determining the location of snowlines, and how well these can be observed and measured.
         
        \begin{acknowledgements}
        	N.M.M.~acknowledges support from the RIKEN Special Postdoctoral Researcher Program (Fellowships).
        	C.W.~acknowledges financial support from the University of Leeds, the Science and Technology Facilities Council, and UK Research and Innovation (grant numbers ST/T000287/1 and MR/T040726/1).
        \end{acknowledgements}
        
        \bibliographystyle{aa}
        \bibliography{snowline_models}
        
        \begin{appendix}
                \section{\ce{CO} isotopologues}
                \label{app:co_isotopologs}
                Section~\ref{subsec:COline} and Figure~\ref{fig:COsnowlineRT} show the robustness of  \ce{N2H+} as a tracer of the \ce{CO} snowline location.
                In this appendix, \ce{N2H+} emission is compared with \ce{CO} isotopologs to further explore the robustness of \ce{N2H+} as a \ce{CO} snowline tracer.
                The \ce{^{13}CO} and \ce{C^{18}O} emission are less extended than the \ce{^{12}CO} emission, which is expected from observations.
                As the column densities of \ce{^{13}CO} and \ce{C^{18}O} decrease with radius, the optical depth of the lines also decrease, and thus the emission strength declines sharply as well. Hence, r(\ce{^{12}CO}) $>>$ r(\ce{^{13}CO}) $>>$ r(\ce{C^{18}O}), where r($X$) is the snowline location of $X$ species. It is possible, as well, that \ce{C^{18}O} is optically thin throughout the observed region, whereas \ce{^{12}CO} and \ce{^{13}CO} will remain optically thick out to some distance.

                \begin{figure*}
                        \centering
                        \includegraphics[width=\linewidth]{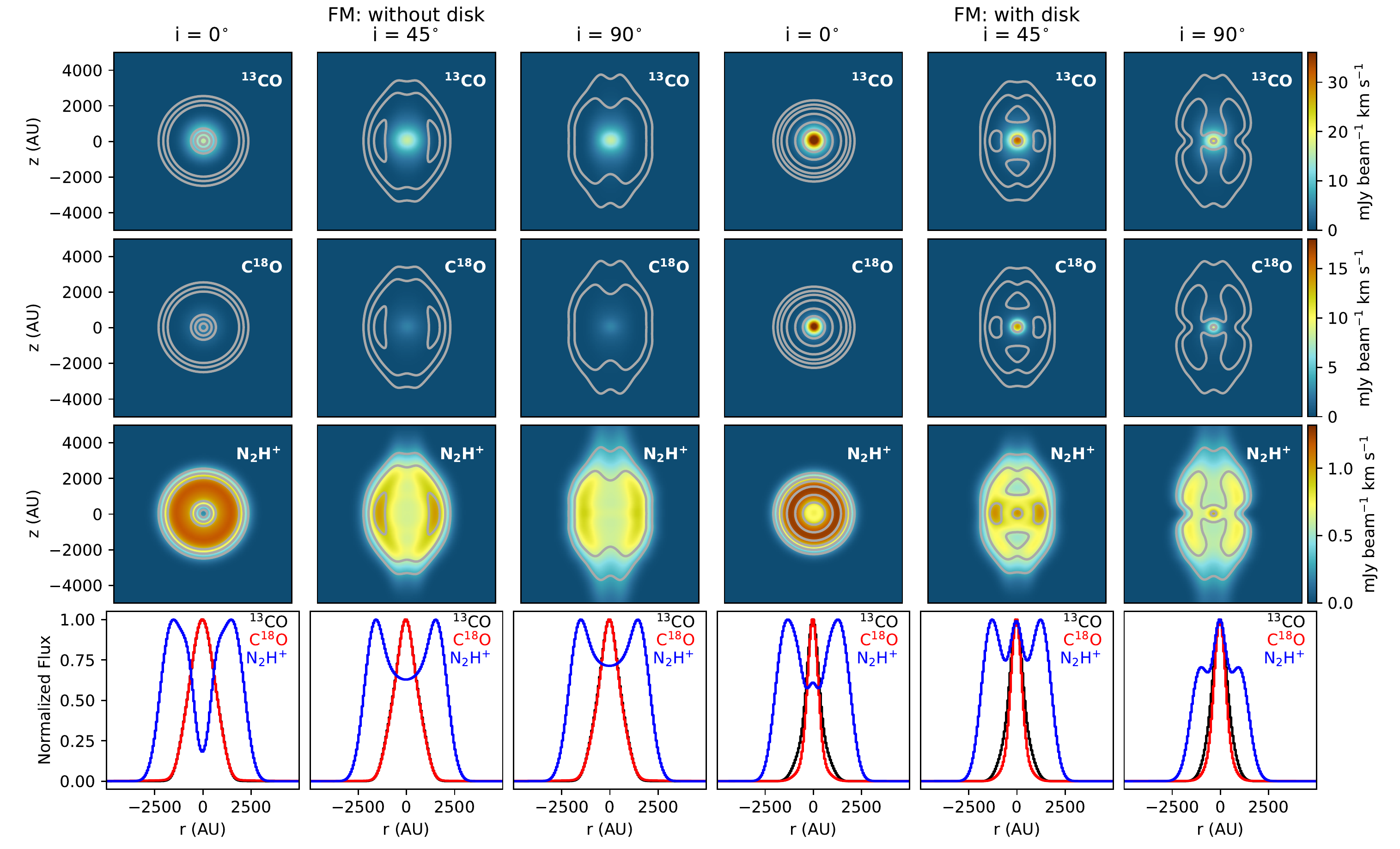}
                        \caption{Robustness of \ce{N2H+} as a \ce{CO} snowline location tracer compared to \ce{CO} isotopolog emission. Simulated emission maps show \ce{^{13}CO} (\textit{top row}), \ce{C^{18}O} (\textit{second row}) and \ce{N2H+} (\textit{third row}) emission. Contours in all three rows are \ce{N2H+} at 0.30, 0.6, 0.9, 1.2, 1.5 mJy~beam$^{-1}$ km~s$^{-1}$. Different inclinations are shown in order to determine if inclination affects the robustness of \ce{N2H+} as a \ce{CO} snowline position tracers. Inclinations shown are from face-on ($i = 0^{\circ}$) to edge-on ($i = 90^{\circ}$). The first three columns show the fiducial model without disk, while the other three columns show the fiducial model with disk. Slices extracted along $z = 0$ from the simulated line emission maps are shown in the bottom row normalized to the peak of each profile.}
                        \label{fig:COiso_snowlineRT}
                \end{figure*}
                
                \section{Sample of model grid}
                \label{app:models}
                The full model grid is available online.
                The model grid includes the calculated molecular distributions of the species included in the chemical network used in this work (Table~\ref{tab:chemnetwork}).
                The molecular distributions are provided as ascii tables.
                Simulated emission maps of \ce{N2H+} and \ce{HCO+} in fits format are also included for seven inclinations: $i =$ 0$^{\circ}$, 15$^{\circ}$, 25$^{\circ}$, 45$^{\circ}$, 65$^{\circ}$, 75$^{\circ}$, and 90$^{\circ}$.
                In addition, the spectral energy distribution for each inclination is provided.
                Finally, the peak positions of \ce{N2H+} and \ce{HCO+} versus luminosity are provided for each set of models with the same parameters. These are the data used to generate the plots in Figures~\ref{fig:LcenLbol_vspeak} to \ref{fig:disk_vspeak}.
                
\begin{figure*}
                        \centering
                        \includegraphics[width=\linewidth]{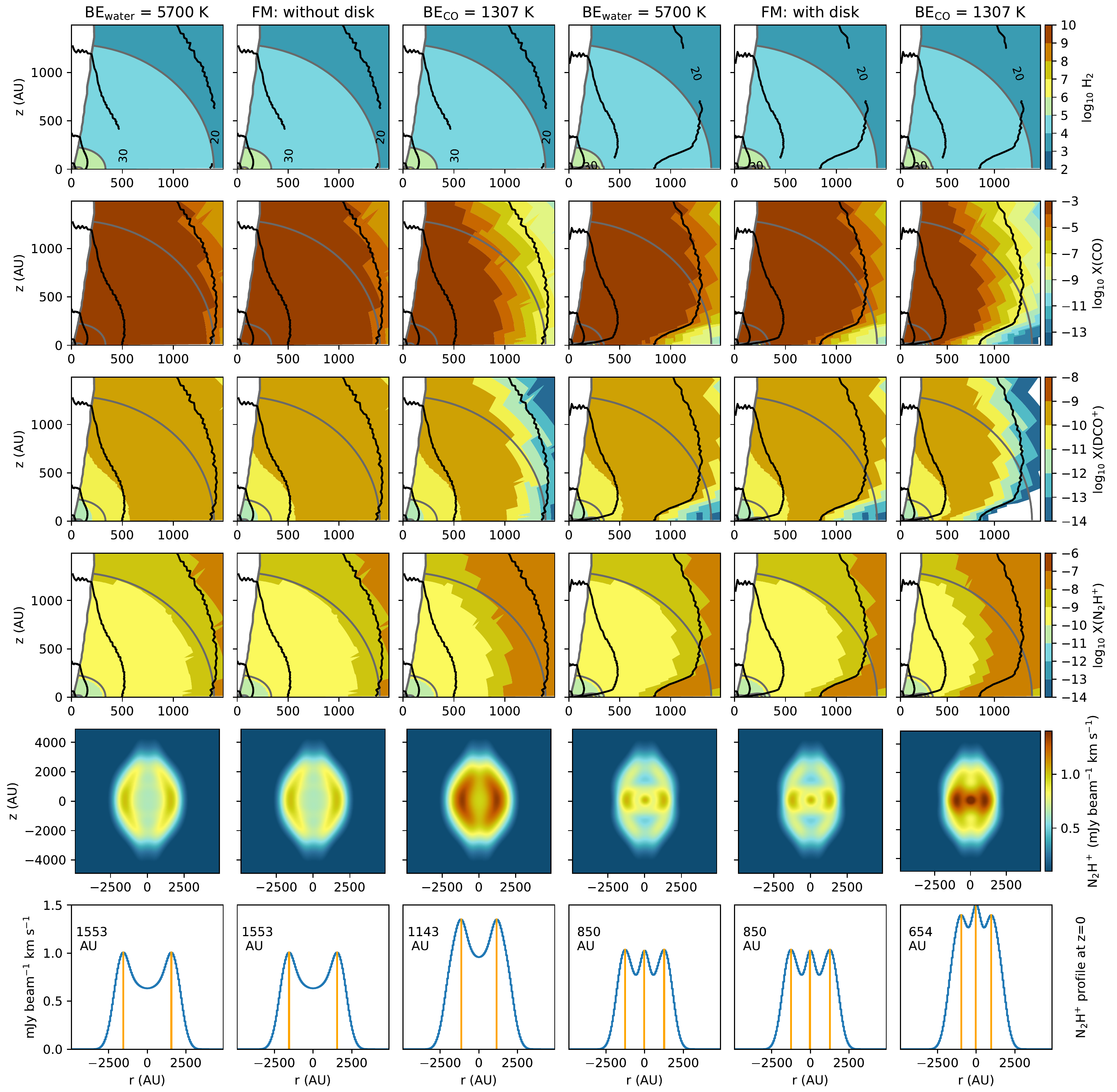}
                        \caption{Effect of binding energy on the \ce{CO} snowline location. Molecular distributions are shown in the top four rows. The fifth row shows the intensity integrated simulated line emission maps for \ce{N2H+} 1--0. The sixth row shows the corresponding slice extracted along $z = 0$ from the \ce{N2H+} simulated emission maps. The average distance of the peaks from the center (i.e., peak radius) is indicated in AU in the top left corner. These positions are shown with orange vertical lines. The second, and fifth columns show the fiducial models without, and with disk, respectively, with $BE_{\rm \ce{CO}}$ = 1150 K and $BE_{\rm \ce{H2O}}$ = 4820 K. The column to the left of each fiducial model has $BE_{\rm \ce{H2O}}$ = 5700 K, while the right column has $BE_{\rm \ce{CO}}$ = 1307 K. Fractional abundances of all molecular species are relative to total number density of \ce{H2} (\textit{top row}). The black and gray contours show gas temperature and density, respectively. The simulated line emission maps are shown at $i$ = 45$^{\circ}$ and convolved to a beam of 2$\arcsec$.} 
                        \label{fig:BE}
                \end{figure*}

\begin{figure*}
                        \centering
                        \includegraphics[width=\linewidth]{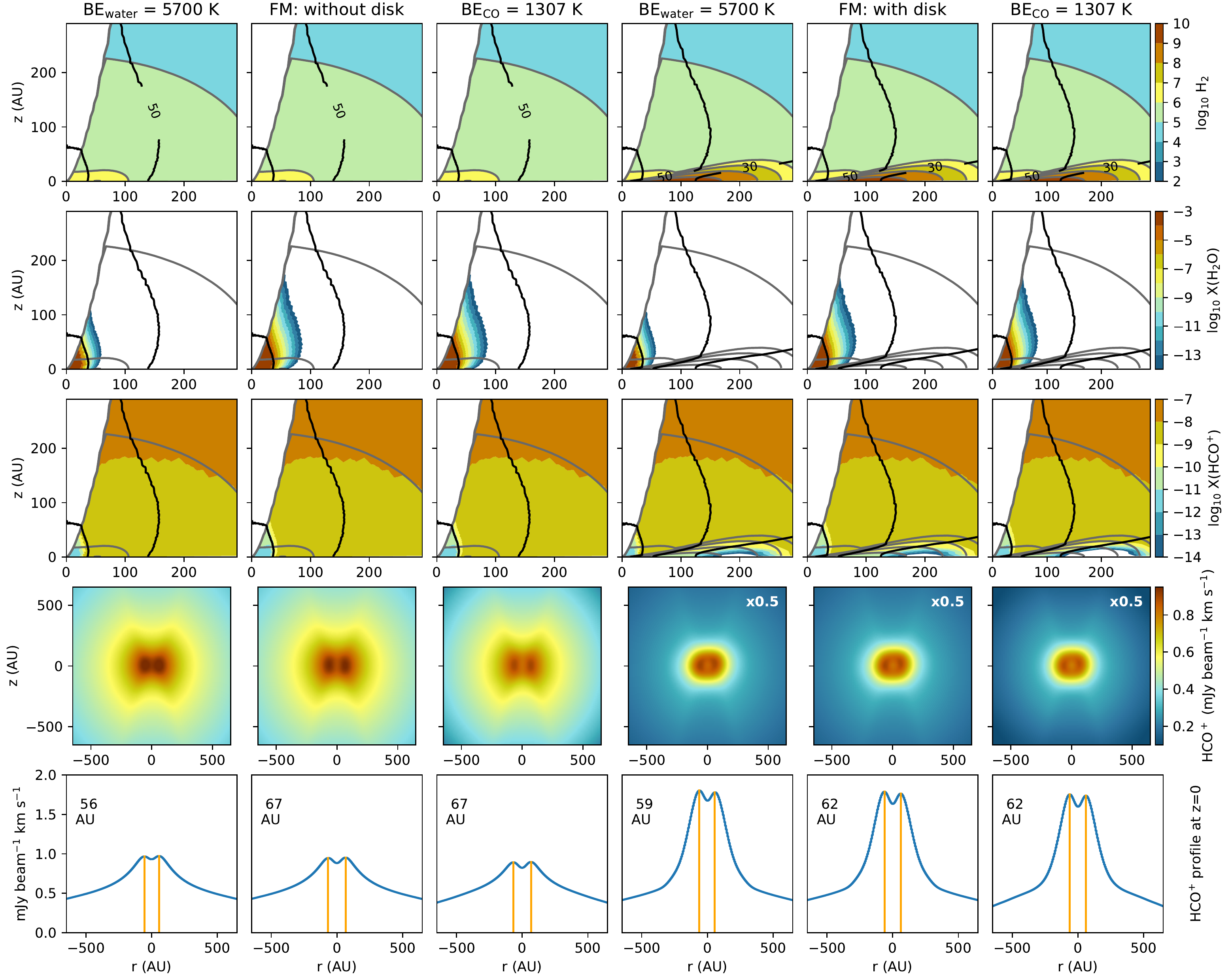}
                        \caption{Effect of binding energy on the \ce{H2O} snowline location. Molecular distributions are shown in the top three rows. The fourth row shows the intensity integrated simulated line emission maps for \ce{HCO+} 3--2. Images that have been scaled for better comparison have the scaling factor on the top right corner. The fifth row shows the corresponding slice extracted along $z = 0$ from the \ce{HCO+} simulated emission maps. The average distance of the peaks from the center (i.e., peak radius) is indicated in AU in the top left corner. These positions are shown with orange vertical lines. No scaling has been applied to the \ce{HCO+} profiles. The second, and fifth columns show the fiducial models without, and with disk, respectively, with $BE_{\rm \ce{CO}}$ = 1150 K and $BE_{\rm \ce{H2O}}$ = 4820 K. The column to the left of each fiducial model has $BE_{\rm \ce{H2O}}$ = 5700 K, while the right column has $BE_{\rm \ce{CO}}$ = 1307 K. Fractional abundances of all molecular species are relative to total number density of \ce{H2} (\textit{top row}). The black and gray contours show gas temperature and density, respectively. The simulated line emission maps are shown at $i$ = 45$^{\circ}$ and convolved to a beam of 0.25$\arcsec$.}
                        \label{fig:BEH2O}
                \end{figure*}

\begin{figure*}
                \centering
                \includegraphics[width=\linewidth]{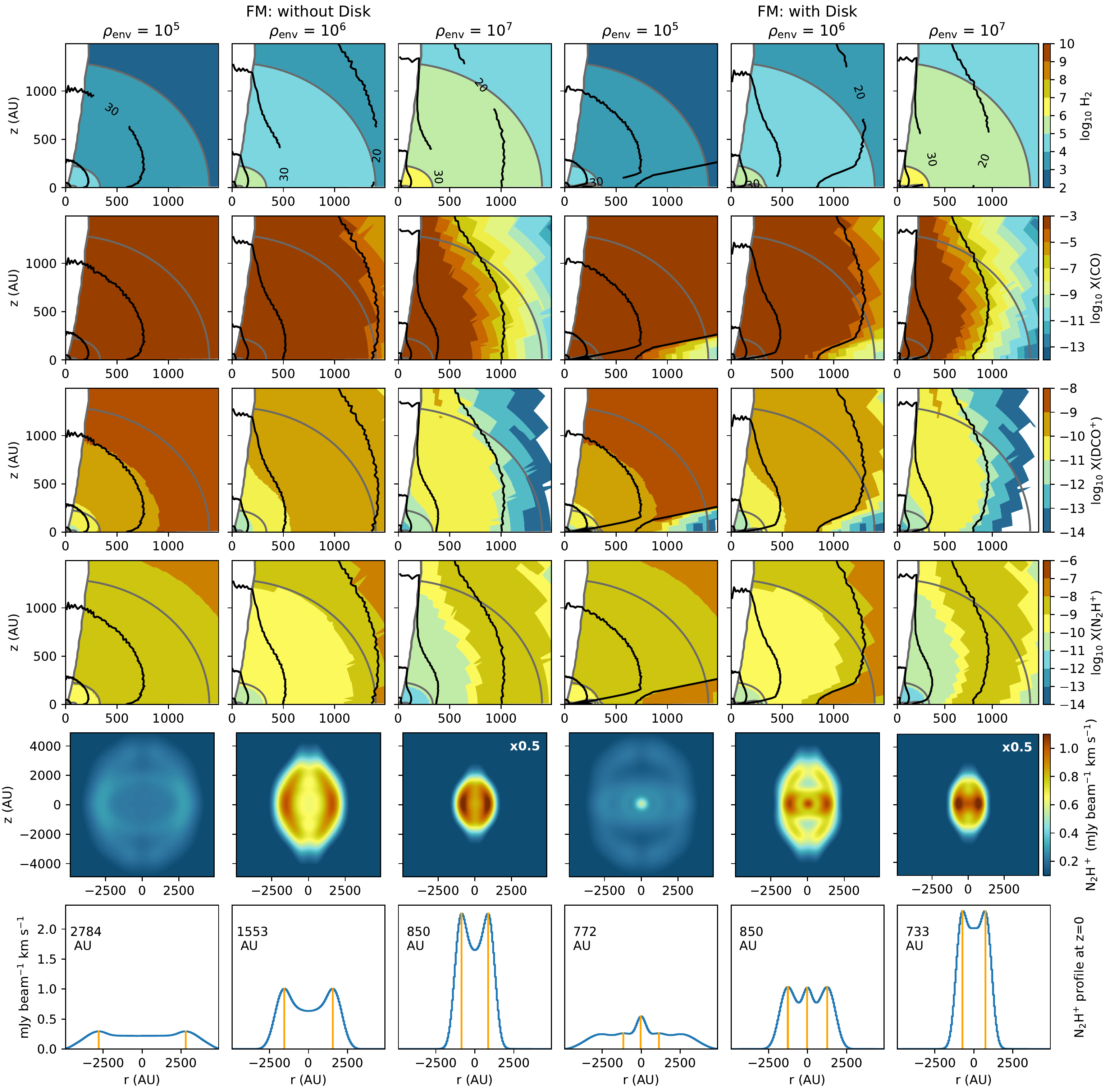}
                \caption{Effect of envelope density on the \ce{CO} snowline location. Molecular distributions are shown in the top four rows. The fifth row shows the intensity integrated simulated line emission maps for \ce{N2H+} 1--0. Images that have been scaled for better comparison have the scaling factor on the top right corner. The sixth row shows the corresponding slice extracted along $z = 0$ from the \ce{N2H+} simulated emission maps. The average distance of the peaks from the center (i.e., peak radius) is indicated in AU in the top left corner. These positions are shown with orange vertical lines. No scaling has been applied to the \ce{N2H+} profiles. The second and fifth columns show the fiducial model without and with disk, respectively, having a density of $\rho_{\rm env}$ = 10$^{6}$ cm$^{-3}$. First and fourth columns show densities of $\rho_{\rm env}$ = 10$^{5}$ cm$^{-3}$, while the third and sixth columns show $\rho_{\rm env}$ = 10$^{7}$ cm$^{-3}$. Fractional abundances of all molecular species are relative to total number density of \ce{H2} (\textit{top row}). The black and gray contours show gas temperature and density, respectively. The simulated line emission maps are shown at $i$ = 45$^{\circ}$ and convolved to a beam of 2$\arcsec$.}
                \label{fig:density}
            \end{figure*}

\begin{figure*}
                        \centering
                        \includegraphics[width=\linewidth]{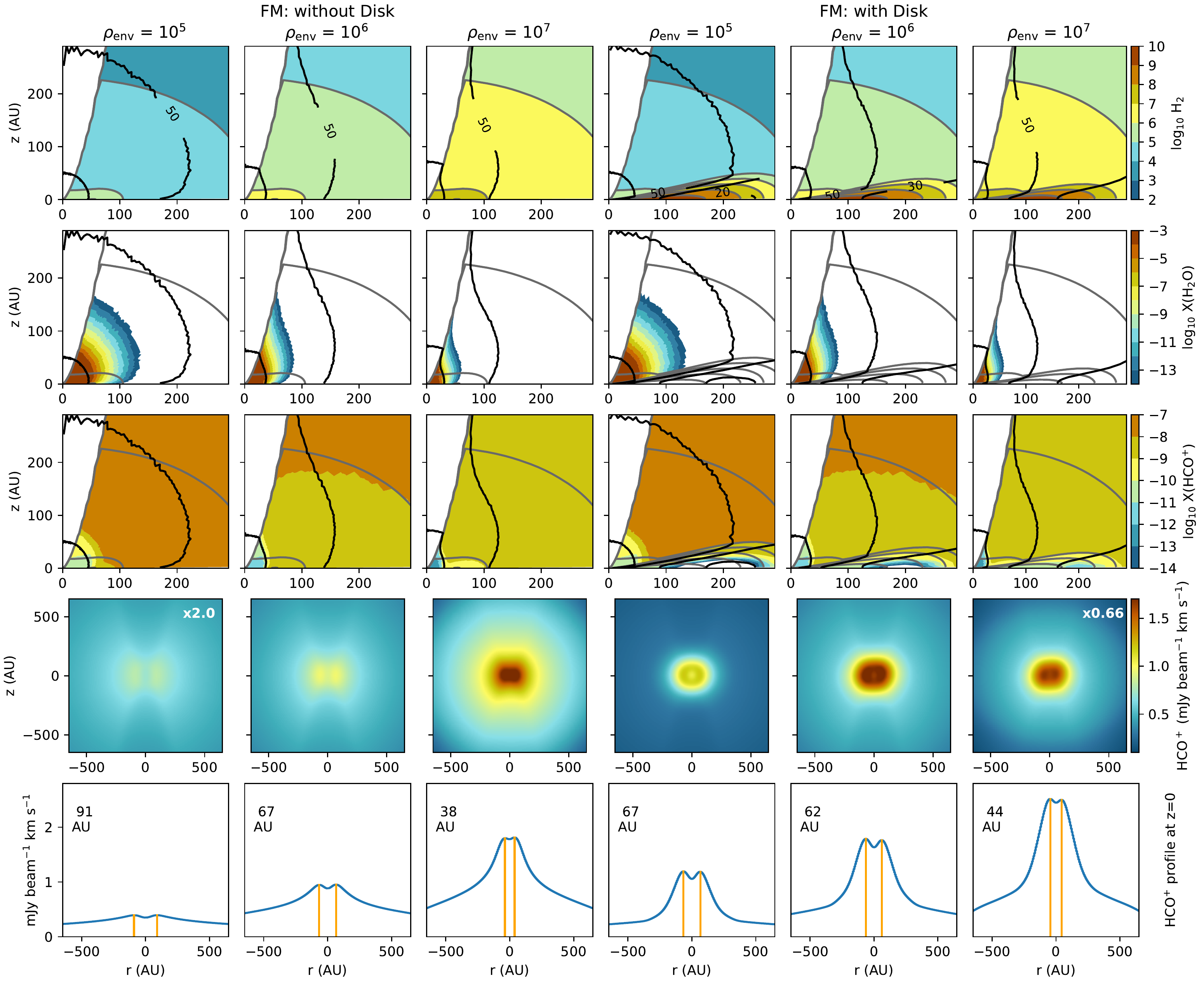}
                        \caption{Effect of envelope density on the \ce{H2O} snowline location. Molecular distributions are shown in the top three rows. The fourth row shows the intensity integrated simulated line emission maps for \ce{HCO+} 3--2. Images that have been scaled for better comparison have the scaling factor on the top right corner. The fifth row shows the corresponding slice extracted along $z = 0$ from the \ce{HCO+} simulated emission maps. The average distance of the peaks from the center (i.e., peak radius) is indicated in AU in the top left corner. These positions are shown with orange vertical lines. No scaling has been applied to the \ce{HCO+} profiles. The second and fifth columns show the fiducial model without and with disk, respectively, having a density of $\rho_{\rm env}$ = 10$^{6}$ cm$^{-3}$. First and fourth columns show densities of $\rho_{\rm env}$ = 10$^{5}$ cm$^{-3}$, while the third and sixth columns show $\rho_{\rm env}$ = 10$^{7}$ cm$^{-3}$. Fractional abundances of all molecular species are relative to total number density of \ce{H2} (\textit{top row}). The black and gray contours show gas temperature and density, respectively. The simulated line emission maps are shown at $i$ = 45$^{\circ}$ and convolved to a beam of 0.25$\arcsec$.}
                        \label{fig:densityH2O}
                \end{figure*}

\begin{figure*}
                \centering
                \includegraphics[width=0.75\linewidth]{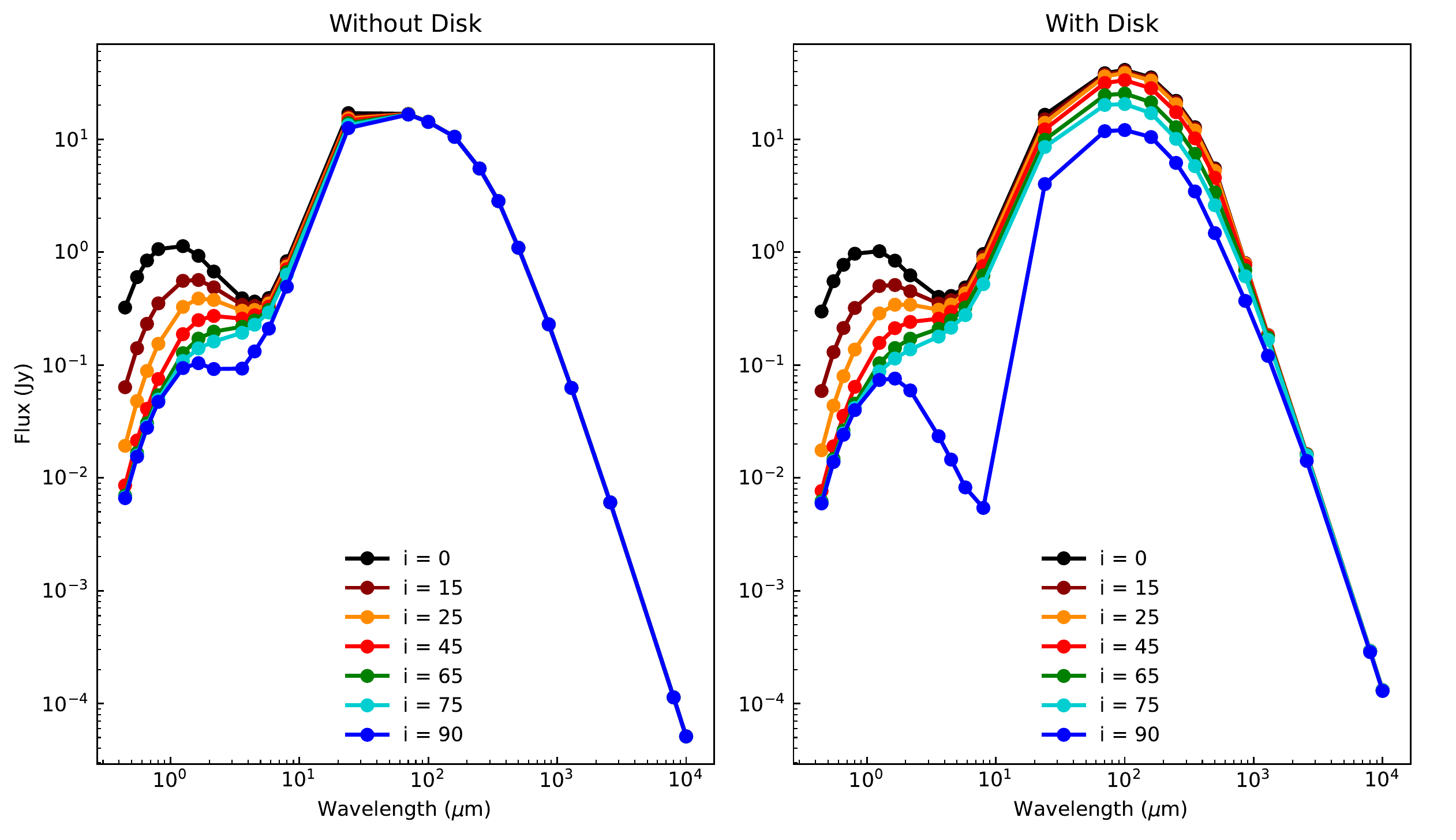}
                \caption{Calculated spectral energy distributions (SEDs) for the fiducial models, without disk (\textit{left}) and with disk (\textit{right}). The different colored lines indicate inclination.}
                \label{fig:SED}
            \end{figure*}

\begin{figure*}
                \centering
                \includegraphics[width=0.8\linewidth]{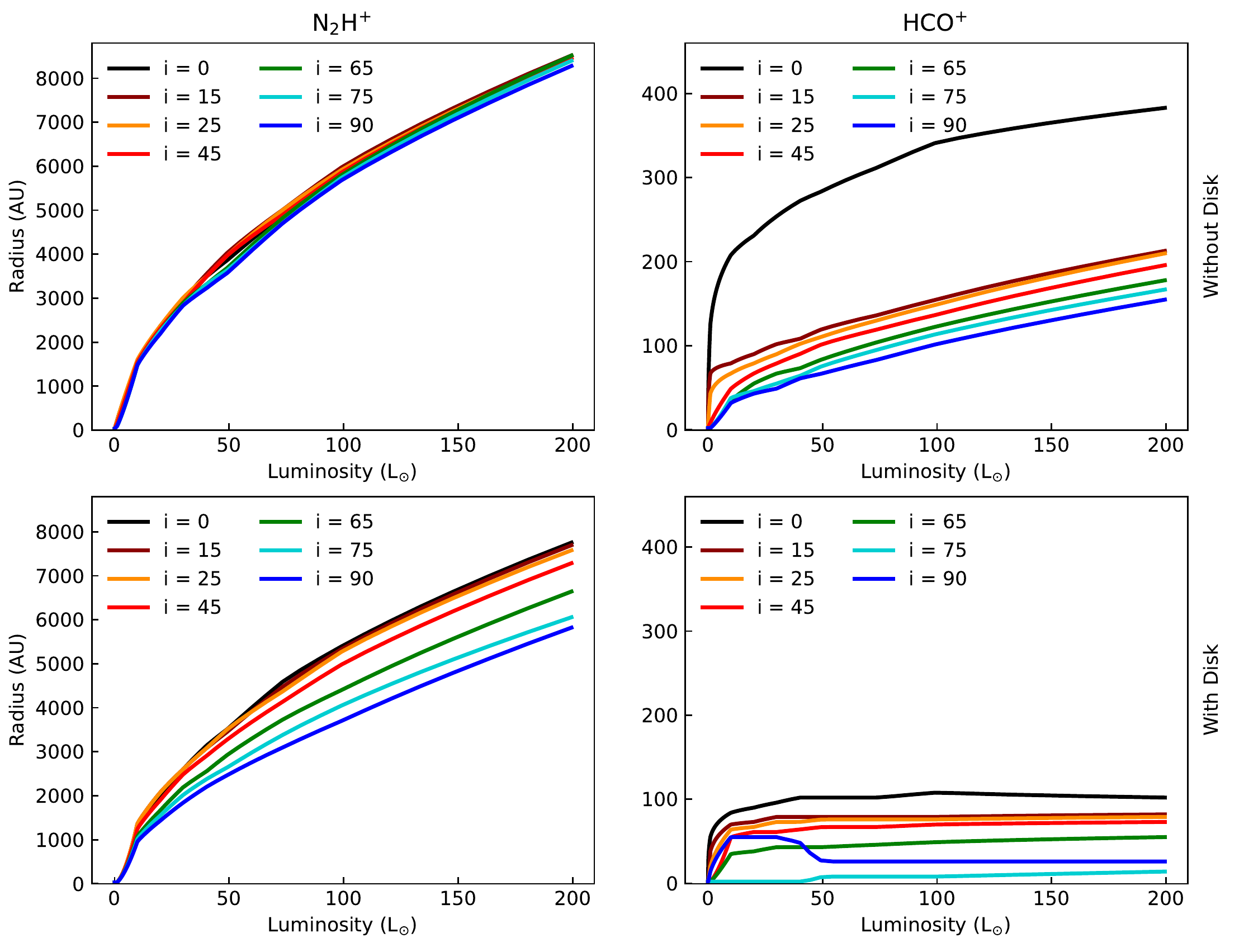}
                \caption{Peak position of \ce{N2H+} (\textit{left}) and \ce{HCO+} (\textit{right}) versus luminosity for the disk, envelope and outflow cavity conditions of the fiducial models. The different colors indicate inclination. These curves are combined into the black curve shown in Figures~\ref{fig:LcenLbol_vspeak} to \ref{fig:disk_vspeak}. We note that the peak position of \ce{N2H+} does not show a clear trend with inclination, while the peak position of \ce{HCO+} decreases with increasing $i$, highlighting the importance of inclination in measuring an accurate water snowline location.}
                \label{fig:inclination}
        \end{figure*}

                In this appendix a sample of the data available from the model grid is shown.
                Figures~\ref{fig:BE} to \ref{fig:densityH2O} show comparisons of the models without and with disk for changes in binding energy and envelope density.
                These models are discussed in the main text.
                Figure~\ref{fig:SED} shows the spectral energy distributions (SEDs) for all inclinations used in this work for the two fiducial models.
            It is interesting to note that the presence of the disk changes the brightness of the SED peak with increasing inclination.
        Figure~\ref{fig:inclination} illustrates how the cloud core inclination affects the measured peak position of \ce{N2H+} and \ce{HCO+} with respect to luminosity for the disk, envelope and outflow cavity conditions of the fiducial models. The models presented in Figure~\ref{fig:inclination} are the same as the black curves in Figures~\ref{fig:LcenLbol_vspeak} to \ref{fig:disk_vspeak}.
                
        \end{appendix}
        
\end{document}